\documentclass[12pt,preprint]{aastex}
\usepackage{natbib}
\usepackage{xcolor}

\shorttitle{}
\shortauthors{Hao, Q. et al.}

\begin{document}

\title{Statistical Analysis of Filament Features Based on the H$\alpha$ Solar Images from 1988 to 2013 by Computer Automated Detection Method}

\author{Q. Hao$^{1,2,3,4}$, C. Fang$^{1, 2}$, W. Cao$^{3, 4}$, P. F. Chen$^{1, 2}$}

\affil{$^1$School of Astronomy and Space Science, Nanjing
University, Nanjing 210023, China}

\affil{$^2$Key Laboratory of Modern Astronomy and Astrophysics, Ministry of Education, Nanjing, China}

\affil{$^3$New Jersey Institute of Technology, Center for Solar Research, 323 Martin Luther King Blvd., Newark, NJ 07102, USA}

\affil{$^4$Big Bear Solar Observatory, 40386 North Shore Lane, Big Bear City, CA92314-9672, USA}

\email{fangc@nju.edu.cn, haoqi@nju.edu.cn}

\begin{abstract}
We improve our filament automated detection method which was proposed in our previous works. It is then applied to process the full disk H$\alpha$ data mainly obtained by Big Bear Solar Observatory (BBSO) from 1988 to 2013, spanning nearly 3 solar cycles. The butterfly diagrams of the filaments, showing the information of the filament area, spine length, tilt angle, and the barb number, are obtained. The variations of these features with the calendar year and the latitude band are analyzed. The drift velocities of the filaments in different latitude bands are calculated and studied. We also investigate the north-south (N-S) asymmetries of the filament numbers in total and in each subclass classified according to the filament area, spine length, and tilt angle. The latitudinal distribution of the filament number is found to be bimodal. About $80\%$ of all the filaments have tilt angles within $[0^{\circ}, 60^{\circ}]$. For the filaments within latitudes lower (higher) than $50^{\circ}$ the northeast (northwest) direction is dominant in the northern hemisphere and the southeast (southwest) direction is dominant in the southern hemisphere. The latitudinal migrations of the filaments experience three stages with declining drift velocities in each of solar cycles 22 and 23, and it seems that the drift velocity is faster in shorter solar cycles. Most filaments in latitudes lower (higher) than $50^{\circ}$ migrate toward the equator (polar region). The N-S asymmetry indices indicate that the southern hemisphere is the dominant hemisphere in solar cycle 22 and the northern hemisphere is the dominant one in solar cycle 23.
\end{abstract}

\keywords{ Sun: filaments, prominences -- Sun: chromosphere --Sun: magnetic fields -- techniques: image processing}

\section{Introduction}

\citet{1858Carrington} discovered that the locations of sunspots drift in latitude toward the solar equator within a solar cycle. \citet{1913Maunder,1922Maunder} named the time-latitude plot of sunspots as the well known ``Butterfly diagram". Since then, the latitude migration of solar activity has been studied extensively for a variety of solar activity indices, including solar filaments \cite[e.g.,][]{2010Li}. Besides the sunspot number and sunspot area, solar filaments are also one important indicator of solar activity, which has been considered by many authors \citep{1998Coffey,2008Li,2010Li,2012Gao,2013Hao,2014Zou}.

Solar filaments are prominences projected against the solar disk, which are 100 times cooler and denser ``clouds" suspended in the solar corona. They  are particularly visible in H$\alpha$ observations, where they often appear as elongated dark sticks with several barbs \citep{1995Tandberg}. The filaments are always aligned with photospheric magnetic polarity inversion lines (PIL), and magnetic flux cancellation also often occurs at the photosphere close to it \citep{1998Martin}. Sometimes they undergo hydromagnetic instabilities, which break their equilibria and lead to eruptions. Since filament eruptions are often associated with flares and coronal mass ejections (CMEs) which are the major driving sources of hazardous space weather \citep{2000Gilbert,2003Gopalswamy,2004Jing,2008Chen,2011Chen,2011Shibata,2011Green,2012Webb,2012Zhang}, the understanding of solar filaments becomes crucial in space weather research \citep{2014Chen,2015Low}.

Several features of solar filaments might be meaningful, such as the location, size, tilt angle, and barbs. These features vary over a wide range \citep{1981Hundhausen,2007Heinzel,2008Lin}. Solar filaments are distributed around the whole solar disk, from the solar equator to polar regions during the whole period of each solar cycle. Such a distribution is not random, it actually traces the special magnetic PILs on the solar surface. Hence both the location and the tilt angle provide useful insight into the nature of the solar magnetic field, which is determined by the emergence and diffusion of sunspots \citep{1972McIntosh,1994Mouradian,1998Minarovjech,1998Rusin,2000Rusin}.  The size of a filament might determine the size of the two-ribbon flare after the filaments erupts. Barbs are another important feature since it might disclose the secret of the magnetic configuration of the filament. In order to figure out the magnetic configuration of filaments, several models have been constructed, such as wire model \citep{1994Martin,1998Martin2} and flux rope model \citep{1998Aulanier,1998Aulanier2,1999Aulanier}. Recently, with the help of the relatively high-resolution observations and the available data of vector magnetic filed, \citet{2010Guo1} and \citet{2015Hao} found that the filaments sometimes have a complex magnetic structure which formed partly by a flux rope and partly by a dipped arcade, and the difference can be revealed directly by the orientation of barbs \citep{2010Guo1,2014Chen}.

In order to derive all these information on filaments, automated pattern recognition method should be developed.
Since \citet{2002Gao} combined the intensity threshold and region growing method to develop an algorithm to automatically detect the growth and the disappearance of filaments, a number of automated filament detection using appropriate algorithms have been developed in the past decade \citep{2003Shih,2005Fuller,2005Bernasconi,2005Qu,2010Wang,2010Labrosse,2011Yuan,2013Hao}. Recently, we have improved our filament automated detection method in our previous works \citep{2013Hao}. Besides the algorithms applied in our previous method, we also adopt a cascading Hough circle detector to find and dually check the center location and the radius of the solar disk; use the polynomial surface fitting technique to correct the unbalanced luminance; build an adaptive filter based on the Otsu's method to find an optimum threshold in order to segment filament and non-filament features.

In this paper, we use our improved filament automatic detection method to process full disk H$\alpha$ data mainly obtained by Big Bear Solar Observatory (BBSO) from 1988 to 2013. We focus on analyzing the processed results in order to find statistical characteristics of the filament features. In Section~\ref{methods}, we describe the improvement of our automatic detection algorithm. We give the results of automatically detected filament features in Section~\ref{results}, and our discussion and conclusion are given in Section~\ref{conclusion}.

\section{Methods}
\label{methods}
We described our efficient and versatile automated detecting and tracing algorithm for solar filaments extensively in our previous work \citep{2013Hao}. Here, we only highlight and explain the improvements of our method in the following subsections.

\subsection{Solar Disk Center Location and Radius Determination}
\label{solardisk}
The changing distance between the Earth and the Sun can cause the variation of the solar radius. Besides, H$\alpha$ images obtained by different telescopes may also have different values of the solar radius in units of pixels. Therefore, the first step of the image processing is to determine the solar disk center and the solar radius in the images. In our previous work, we applied the Canny edge detection technique to separate filaments in addition to obtain the limb of the solar disk. We then used the circle fitting algorithms proposed by \citet{1987Pratt} and \citet{1991Taubin} to fit the solar limb and calculate the geometric center and the radius of the fitted circle. Here, we introduce a method called cascading Hough circle detector which was presented by \citet{2011Yuan}, and combine it with our previous method to dually pre-process the data in order to accurately determine and check the center location and the radius of the solar disk. If the difference between the values obtained with the two methods is greater than 2 pixels, the detected results will be treated as failure and the data will be automatically recorded for further manual check. These two methods perform quite well in the test data base, only less than 5$\%$ of all the images need to be further manually checked.

\subsection{Non-uniform Brightness Correction}
\label{luminance}
Since the data used in our previous work were relatively good, we only need to remove the effect of limb-darkening. However, the non-uniform brightness is not only caused by limb-darkening, but also by clouds in the Earth atmosphere or dusts on the telescope lens, and so on. In order to process the data within a long time interval and from various observatories, we adopt a more efficient brightness non-uniformity correcting method, i.e., the polynomial surface fitting technique, which was also applied by \citet{2005Bernasconi} and \citet{2011Yuan}. The detailed description of this method can be found in the appendix of \citet{2011Yuan}. The only modification here is that we process the pixels within the detected solar disk instead of the whole image.

\subsection{Adaptive Threshold Filter for Non-filament Structures}
\label{threshold}
Frequently people use an appropriate global-threshold to filter out the bright features such as plages and flares in order to get the dark features, such as filaments and sunspots in the H$\alpha$ images. This method works well for the data from one observatory. In other words, if the data are obtained through various observatories, the global threshold is not appropriate since the seeing conditions and the image qualities are different from one to another. Therefore, an adaptive threshold filter is needed to process H$\alpha$ data from different observatories. We designed an adaptive threshold filter based on Otsu's method \citep{1979Otsu}. The algorithm assumes that each image contains two classes of pixels or a bi-modal histogram (e.g., foreground and background, here the foreground includes dark features such as filaments and sunspots, whereas the background includes other features), then calculates the optimum threshold separating those two classes so that their combined spread (intra-class variance) is minimal (or, conversely, to maximize the between-class variance). The procedure of the algorithm is detailed as below.

Assuming that the input image is $I(x,y)$ which is represented in $L$ grayscale levels. The number of pixels at level $i$ is denoted by $n_{i}$ and the total number of the pixels it contains is $N=n_{1}+n_{2}+\cdots+n_{L}$. The probability of the $i$th grayscale level is $p_{i}=n_{i}/N$ ($p_{i}\geq0, \sum^{L}_{i=1}p_{i}=1$). The pixels are dichotomized into two classes, i.e., background and object (filaments here), by a threshold at level $k$. In Otsu's method we exhaustively search for the threshold that maximizes the between-class variance:
\begin{equation}\label{eq1}
	\sigma^{2}=\omega_{1}(\mu_{1}-\mu_{T})^{2}+\omega_{2}(\mu_{2}-\mu_{T})^{2}
	=\omega_{1}\omega_{2}(\mu_{1}-\mu_{2})^{2} ,
\end{equation}
where  $\omega_{1}=\sum^{k}_{i=1}p_{i}$ and $\omega_{2}=\sum^{L}_{i=k+1}p_{i}$ are the probability of each class, $\mu_{1}=\sum^{k}_{i=1}ip_{i}/\omega_{1}$ and $\mu_{2}=\sum^{L}_{i=k+1}ip_{i}/\omega_{2}$ are the weighted mean level of each class, and $\mu_{T}=\sum^{L}_{i=1}ip_{i}$ is the weighted mean level of the whole original image $I(x,y)$. Formula (\ref{eq1}) is the original Otsu algorithm, the following formula is the modified Otsu algorithm that we used:
\begin{equation}\label{eq1.2}
	\sigma^{*2}=\omega_{1}\omega_{2}(\mu_{1}/\mu_{T}-\mu_{2}/\mu_{T})^{2} 
	\approx \omega_{1}\omega_{2}(\omega_{1}-\omega_{2})^{2} ,
\end{equation}
Compared to formula (\ref{eq1}), formula (\ref{eq1.2}) avoids the decimal rounding errors resulting from the multiplication between the pixel grayscale and its probability, and also has a higher efficiency. Then, the optimal threshold $k^{*}$ is taken so that
\begin{equation}\label{eq2}
	\sigma^{*2}(k^{*})= \max \limits_{1 \leq k \leq L} \sigma^{*2}(k).
\end{equation}
We scan through all possible thresholds ($1, 2, \cdots, L$) and obtain the threshold $T_{Otsu}=k^{*}$ which makes $\sigma^{2}(k)$ ($1 \leq k \leq L$) maximal. The value of $k^*$ will be treated as the threshold to filter out non-filament structures. It is noticed that filaments occupy only a small part of the solar disk, therefore, the number of corresponding pixels is very small. When applying Otsu's method, we found that the resulting threshold is usually higher than the proper threshold so that the non-filament structures are not filtered out accurately. To overcome this shortcoming, here we introduce the mean grayscale level of the pixels within solar disk as $T_{mean}$ to adjust the threshold:
\begin{equation}\label{eq3}
  T_{final}=\left\{
   \begin{array}{l}
   T_{Otsu}, |T_{Otsu}-T_{mean}| \leq 10,  \\
   T_{Otsu}-\alpha|T_{Otsu}-T_{mean}|,  10 < |T_{Otsu}-T_{mean}| \leq 20   \\
   T_{Otsu}-\beta|T_{Otsu}-T_{mean}|,  |T_{Otsu}-T_{mean}| > 20   \\
   \end{array}
  \right.
  \end{equation}
$T_{final}$ is the final threshold we use for the adaptive threshold to filter out the non-filament structures. $\alpha$ and $\beta$ are the adjustment coefficients. We tested several hundred images that obtained by various observatories at different times and found the appropriate values for them to be $\alpha=1.6$ and $\beta=1.2$.

\subsection{Filament Feature Identification}
\label{feature}
Our method can identify several filament features, such as the location, perimeter, area, spine, filament direction, tilt angle, and barbs. For the filament spines and barbs, we introduce the unweighted undirected graph concept and adopt Dijkstra shortest-path algorithm \citep{1959Dijkstra} to recognize them. We then employ the connected components labeling method to identify the barbs and calculate the angle between each barb and the spine to indicate the bearing sense of the barb. The corresponding algorithms are described in \citet{2015Hao}. Since some data in early time have a relatively low resolution, in this study we only detect the filament spines and the barb number of barbs.

\subsection{Merging Filament Fragment Consolidation}
\label{fragment}
A filament may be split into several fragments during its evolution. We adopted a distance criterion to find the fragments belonging to a single filament \citep{2013Hao}. However, this method has its shortages. For example, sometimes several filaments in an active region are so close to each other that they satisfy the distance criterion. In this case, they would be recognized as one filament. To avoid this error, we add a new requirement, i.e., the tilt angles of the fragments. If the fragments satisfy the distance criterion, meanwhile, they have similar tilt angles, they are treated as one filament. In addition, we compare two footpoints of each filament for the distance criterion instead of the geometric centers of individual fragments as used in our previous work. These changes make the filament fragment consolidation more accurate.

\subsection{Data Acquisition and Method Implementation}
\label{data}
We tested the performance of our improved method with the observations from various observatories, including the Optical \& Near Infrared Solar Eruption Tracer (ONSET) in Nanjing University \citep{2013Fang}, Mauna Loa Solar Observatory (MLSO), Big Bear Solar Observatory (BBSO), Kanzelh\"{o}he Solar Observatory (KANZ), Catania Astro physical Observatory (OACT), and Yunnan Astronomical Observatory (YNAO). The test results show that our improved method is much more efficient and universal. For example, although the BBSO H$\alpha$ archive data observed in different periods have different resolutions, say, $640 \times 481$, $1016 \times 1014$, $1844 \times 1844$, $1903 \times 1903$, $1928 \times 1928$,  $1981 \times 1981$, $2032 \times 2032$ and $2048 \times 2048$ pixels, our method can still handle these variations efficiently. In this study, we use our method to process one image per day of all available H$\alpha$ data in the BBSO H$\alpha$ archive (\url{ftp://ftp.bbso.njit.edu/pub/archive}). The available data are from 1988 to 2014. Note that their data between 2005 and 2007 are rare. Thanks to the Global High Resolution H$\alpha$ Network, data from other observatories can fill in the gap. In order to maintain the data consistency we use the data from KANZ.

Our detection method for the filament features is developed by using the MATLAB Desktop Tools and Development. In addition to using MATLAB to implement the algorithm for the filament detection, we also use Interactive Data Language (IDL) with the solar software (SSW) library to pre-process the raw data. The time consumption of these methods is about several seconds for one image.

\section{Results}
\label{results}
We processed 6564 images mainly obtained by BBSO from April 1988 to December 2013 with our improved automated filament detection method.  In this section, we present the statistical results of the filament features in the following subsections.

\subsection{Characteristics of Various Filament Features}
\label{distribution}

\subsubsection{Filament Number Distribution}
\label{Fnumber}

In order to see the difference of the filament features during the three solar cycles, we divide the observations into 3 intervals: from April 1988 to August 1996 within solar cycle 22, from September 1996 to December 2008 within solar cycle 23, from January 2009 to December 2013 representing the rising phase of solar cycle 24. Since the normal solar activity is usually applied to mean the events with latitudes lower than $50^{\circ}$ \citep{1998Sakurai,2008Li}, we divide the filaments with latitudes lower (higher) than $50^{\circ}$ in both hemispheres as low (high) latitude filaments for further analysis. The numbers of the processed images and those of the detected filaments are summarized in Table~\ref{table1}. The latitudinal distribution of the filaments in the three solar cycles is plotted in Figure~\ref{fig1}. Panels (a--c) of Figure~\ref{fig1} show the results in solar cycles 22, 23, and the rising phase of solar cycle 24, respectively. The number of filaments in the southern hemisphere is $1,345$ more than that in the northern hemisphere in solar cycle 22, which is opposite in solar cycle 23 and the rising phase of solar cycle 24, during which the detected filament numbers in the northern hemisphere are $994$ and $623$ more than those in the southern hemisphere, respectively. The difference between the filament numbers in both hemispheres is less than $15\%$. The latitudinal distribution of the filament numbers is bimodal. The peak values are within the latitude band [$10^{\circ}$, $30^{\circ}$] in both hemispheres for the three solar cycles, as indicated by Figure~\ref{fig1}.

\subsubsection{Filament Butterfly Diagrams}
\label{butterfly}
We detected $67,432$ filaments from $6,564$ images during the period from April 1988 to December 2013. The temporal evolution of the latitudinal  distribution of these filaments is depicted as the scatter plot in Figures~\ref{fig2}--\ref{fig4}, where we can see butterfly diagrams that are similar to that for sunspots. We choose the geometric center of each filament as its location. Each dot represents a single observation. There are several data gaps in the butterfly diagrams since the lack of observations in these periods. In Figure~\ref{fig2}, we use the magnetogram synoptic maps  from National Solar Observatory (\url{ftp://vso.nso.edu/synoptic/level3/vsm/merged/carr-rot/}) and \textit{Global Oscillation Network Group} (\textit{GONG}, \url{http://gong.nso.edu/}) to draw the latitudinal evolution of the line-of-sight magnetic field from the year 1998 to 2014, which is shown as color scale where the locations of the filaments are superposed as green dots. Besides the well-known butterfly pattern manifested in the magnetic field diagram, we see a similar butterfly diagram in the filament distribution as well. As sunspots migrate from middle latitude toward the equator, solar filaments migrate to the equator as well. However, it is seen that filaments are distributed more dispersively than sunspots. The magnetic fields of the following-polarity sunspots migrate toward the poles where they eventually reverse the polar magnetic field around each solar maximum, as indicated by the blue and yellow stripes above the latitude of $40^{\circ}$. Correspondingly, we can see such poleward stripes in the filament distribution as well, especially from the beginning to the maximum in both solar cycles 22 and 23. These diagrams indicate that filaments mainly migrate toward the solar equator from the beginning to the end of each solar cycle. The colors indicate the filament areas in Figure~\ref{fig3} and the spine lengths in Figure~\ref{fig4}. We can see from Figure~\ref{fig3} that the dots in the period from 1988 to 1995 (namely solar cycle 22) have more deep orange colors than those in the periods from 1997 to 2008 (namely solar cycle 23) and from 2010 to 2013 (within solar cycle 24). It means that the filaments in solar cycle 22 have relatively larger areas. However, it may be because that the data from 1988 to 1995 have relatively low spatial resolution. Similarly, Figure~\ref{fig4} reveals that the dots in solar cycle 23 have more deep blue colors than those in solar cycles 22 and 24, which means that the filaments in solar cycle 23 have relatively longer spines. Therefore, we do not present the results of filament areas, spine lengths and barb numbers from 1988 to 1995, i.e., the period during solar cycle 22 in the following filament feature analysis. 

\subsubsection{Distribution of Filament Area}
\label{area}
In this subsection, we arrange the detected filaments into four groups according to their areas, i.e., $<2.5\times10^{8}$ km$^{2}$, $2.5\times10^{8} \sim 5.0\times10^{8}$ km$^{2}$, $5.0\times10^{8} \sim 1.0\times10^{9}$ km$^{2}$, and  $>1.0\times10^{9}$ km$^{2}$. Figure~\ref{fig5} shows the yearly evolution of the number of the filaments in each group. This result is the same as that displayed in Figure~\ref{fig3}. The filament areas are mainly less than $1.0\times10^{9}$ km$^{2}$ from 1996 to 2013. Filaments with areas less than $2.5\times10^8$ km$^{2}$ almost dominate the period from 1996 to 1999 and those in the range of $2.5\times10^{8} \sim 5.0\times10^{8}$ km$^{2}$ nearly dominate the period from 2000 to 2007. It indicates that in the rising phase of solar cycle 23 the filaments are relatively smaller and in the declining phase the filaments become relatively larger. However, the filaments with areas in the range of $2.5\times10^{8} \sim 5.0\times10^{8}$ km$^{2}$ are dominant in the period from 2010 to 2013 in the rising phase of solar cycle 24. 

Panels (a) and (b) of Figure~\ref{fig6} show the latitudinal distributions of the filaments in different area groups during solar cycle 23 and the rising phase of solar cycle 24, respectively. The results indicate that the latitudinal distribution of the filaments in any area range is also bimodal. The histograms reveal that the largest number of filaments appear in the latitude band $10^{\circ}  \sim  30^{\circ}$ in both hemispheres. Filaments with areas less than $1.0\times10^{9}$ km$^{2}$ dominate the latitude band $0^{\circ}  \sim 60^{\circ}$ and those less than  $2.5\times10^{8}$ km$^{2}$ dominate the latitude band $60^{\circ}  \sim 90^{\circ}$. It indicates that the filaments in low latitudes have various sizes. The results of the rising phase of solar cycle 24 are similar to those in solar cycle 23. Table~\ref{table2} lists the quantitative results of the percentage of the filaments in different latitude bands and area ranges. Filaments with areas less than $1.0\times10^{9}$ km$^{2}$  occupy $89.9\%$ and  $89.2\%$ of all the detected filaments in solar cycle 23 and the rising phase of solar cycle 24, respectively.

\subsubsection{Distribution of Filament Spine Length}
\label{spine}

The length of the spine is a significant feature of a filament, which can also indicate the size of the filament. According to the length of the spine, filaments are divided into three groups, namely, $<5.0\times10^{4}$ km, $5.0\times10^{4} \sim 1.0\times10^{5}$ km, and $>1.0\times10^{5}$ km. Figure~\ref{fig7} shows the yearly distributions of the filaments for the three groups. Panels (a) and (b) of Figure~\ref{fig8} show the distributions of the filament spine lengths in each latitude interval during solar cycle 23 and the rising phase of solar cycle 24, respectively. The quantitative results of the percentage of the filaments in different latitude bands and spine length ranges are listed in Table~\ref{table3}. As shown in Figure~\ref{fig7}, the filaments with spine lengths less than $1.0\times10^{5}\ $ km are dominant from 1996 to 2013. Figure~\ref{fig8} shows that the latitudinal distribution of the filaments in any spine length range is also bimodal. The filament number peaks of the bimodal distributions appear in the latitude band $ 10^{\circ}  \sim  30^{\circ}$ in both hemispheres for the two solar cycles. From Table~\ref{table3} we can see that the spine length is mainly less than $1.0\times10^{5}$ km. The filaments in this range occupy $76.6\%$ and $80.4\%$ of the total filaments in solar cycle 23 and the rising phase of solar cycle 24, respectively. This result can also be obtained visually from Figures~\ref{fig7} and \ref{fig8}.

\subsubsection{Distribution of the Filament Tilt Angle}
\label{tiltangle}

The tilt angle is defined as the angle between the fitted filament spine orientation and the local latitude. We set the solar disk center as the coordinate origin so that the disk is divided into four quadrants. We then use a linear polynomial to fit the spine and calculate the slope of the fitted line. The tilt direction (i.e., filament direction) is defined as northwest (northeast) if the slope is positive (negative) in the first or the second quadrant (namely northern hemisphere). Similarly,  the filament direction is southeast (southwest) if the slope is positive (negative) in the third or forth quadrant (namely southern hemisphere). For simplicity, we use NW, NE, SW and SE to refer to the directions northwest, northeast, southwest and southeast, respectively. The tilt angle is arctan of the slope subtract the angle between the local latitude and the solar equator. Its range is [$-90^{\circ}$, $90^{\circ}$]. In contrast, the tilt angle of the sunspot pair inside an active region is defined as the angle between the line connecting the preceding and the following sunspots and the equator. The following sunspot of an active region tends to appear further away from the equator than the preceding sunspot, and the higher the latitude is, the bigger the tilt angle is. This phenomenon is well-known as Joy's law \citep{1919Hale}. Besides, sunspots usually appear at the latitude band below $40^{\circ}$ in both hemispheres. However, filaments are formed not only inside an active region, but also between two neighboring active regions, between an active region and the quiescent region, or inside a decayed active region which differs from its original active region after differential rotation and diffusion \citep{2010Mackay}. As a result, filaments are distributed in much wider latitudes than sunspots, being aligned with the local PILs. These differences lead to that the tilt angles of filaments and sunspots have very different characteristics. It implies that filaments no longer follow Joy's law strictly. Figure~\ref{fig9} shows the yearly evolution of the filament numbers with each of the four tilt directions in both hemispheres. We can see that the northeast (NE) direction is dominant in the northern hemisphere and the southeast (SE) direction is dominant in the southern hemisphere from 1988 to 2013. Panels (a--c) of Figure~\ref{fig10} show the latitudinal distribution of the filament numbers with different tilt directions during solar cycles 22, 23, and the rising phase of solar cycle 24, respectively. In order to see the difference clearly, we use blue color histograms to represent the numbers of filaments with the NW direction in the northern hemisphere and with the SE direction in the southern hemisphere. We also use green color histograms to represent the filament numbers with the NE direction in the northern hemisphere and with the SW direction in the southern hemisphere. For the filaments with latitudes within $ 0^{\circ}  \sim  50^{\circ}$ (namely low latitude filaments) in both hemispheres, NE is the dominant direction in the northern hemisphere and SE is the dominant direction in the southern hemisphere.  For the filaments with latitudes within $ 50^{\circ}  \sim  90^{\circ}$ (namely high latitude filaments) in both hemispheres, NW is the dominant direction in the northern hemisphere and SW is the dominant direction in the southern hemisphere. Table~\ref{table4} lists the quantitative results of the filament number percentage distributions in different latitude bands with different tilt angle directions. Figure~\ref{fig11} show the distributions of the filament tilt angle in the northern and southern hemispheres from 1988 to 2013.  Figure~\ref{fig12}(a), (b) and (c) show the latitudinal distributions of the filament tilt angles within solar cycles 22, 23, and the rising phase of solar cycle 24, respectively. Table~\ref{table5} lists the quantitative results of the percentage of the filaments with different tilt angles in different latitude bands. About $85\%$ of all the filaments have tilt angles within $[0^{\circ}, 60^{\circ}]$ in both hemispheres.

\subsubsection{Distribution of Number of the Filament Barbs}
\label{barb}

Each barb of a filament tends to be broader near the filament spine and converges toward the end point. Martin and her colleagues \citep{1994Martin, 1998Martin2} gave reasonable classifications for the barb bearing: when viewed from either footpoint of a filament, if a barb veers from the filament spine to the right (left), the barb is defined as right-bearing (left-bearing). Although our programs can determine the barb bearing, here we only detect how many barbs each filament has, due to the relatively low resolution data. Figure~\ref{fig13} shows the yearly variation of the number of the filaments with different numbers of barbs from 1996 to 2013. Based on how many barbs each filament has, the filaments are arranged into three groups:  $0 \sim 5$, $6 \sim 10$, and $>$10. Panels (a) and (b) of Figure~\ref{fig14} show the latitudinal distributions of the number of filaments in each group during solar cycle 23 and the rising phase of solar cycle 24, respectively. Table~\ref{table5} lists the percentage of the filaments with different numbers of barbs in different latitude bands. It is seen that about $75\%$ filaments have less than 5 barbs.

\subsection{Latitude Migration and Drift Velocity of Filaments}
\label{drift}

We use the monthly mean latitude of the filaments in the northern and southern hemispheres to analyze the latitudinal migration and  the drift velocity of the filaments. Here, the positive/negative velocity means that the filament migrates toward the polar region/equator in both hemispheres. As mentioned above, we divide the filaments into two groups, namely, low latitude and high latitude ones which are separated at $50^{\circ}$ in both hemispheres. Then we use the cubic polynomial function to fit the monthly mean latitude of the filaments and obtain the drift velocities.

The distributions of the monthly mean latitudes of the filaments in each hemisphere are plotted in Figure~\ref{fig15}. It can be seen that in either hemisphere the evolution of the monthly mean latitude of the filaments has three stages: from the beginning to the maximum of a solar cycle, the drift velocity is relatively higher; after the solar maximum it becomes relatively lower. Across the minimum, the drift velocity becomes divergent and the filaments migrate toward both high and low latitudes. In order to derive the drift velocity, we pick up two time intervals, i.e., from 1988 to 1994 in solar cycle 22, from 1997 to 2004 in solar cycle 23 for the cubic polynomial fitting analysis. Since the drifting of the filaments in periods from 1995 to 1997 and from 2005 to 2009 are divergent, and solar cycle 24 is not finished yet, we do not fit the drift velocities in these three intervals. The results of the cubic polynomial fitting are shown as the red lines in Figure~\ref{fig15}. Since the detected filaments are mostly located in latitudes lower than $50^{\circ}$, the results show that the averaged latitude of the filaments mainly migrates toward the equator. Three panels in Figure~\ref{fig16} show the drift velocities of the filaments in all, low, and high latitudes within solar cycles 22 and 23, where the red lines and the blue lines correspond to the northern and southern hemispheres, respectively. The black dashed lines and the black dotted line show the solar maximum and solar minimum, respectively. The results indicate that most filaments in low latitudes (lower than $50^{\circ}$) migrate toward the equator. Since the filaments within the latitude band $0^{\circ} \sim 50^{\circ}$ occupy $97.7\%$ of all detected filaments, the drift velocities in the latitude band $0^{\circ} \sim 50^{\circ}$ and in all latitudes have similar trends. From Figure~\ref{fig16} we can visually find that from the beginning to the years around the maximum of each solar cycle, the drift velocity is relatively higher; after the solar maximum it becomes relatively lower. As for the filaments in high latitudes, they migrate toward the poles with relatively high velocities during the rising phases of solar cycles 22 and 23. Then, after the maximum they migrate toward the equator. The velocities increase slightly after the maximum, and then decrease later. The drift velocities from the beginning to the maximum have substantially larger variations than those from the maximum to the minimum of these two solar cycles. The drift velocity difference between the two hemispheres is significantly larger in solar cycle 22 compared to that in solar cycle 23.

\subsection{The N-S Asymmetries of Solar Filament Features}
\label{asymmetry}

The occurrence of solar activity features are not uniformly distributed on the solar disk. It has been noticed that more activity features occur in one part of the solar hemisphere than the other during one solar cycle. This phenomenon is known as the north-south (N-S) asymmetry. The N-S asymmetries of several solar activity phenomena, such as sunspot numbers, sunspot groups, sunspot areas, solar flares, prominences (filaments), CMEs, and magnetic fields have been investigated by various authors \citep{2009Li1,2009Li2,2010Bankoti,2004Joshi,2009Joshi,1996Duchlev,2003Li,2007Gao,2014Obridko}. Here, we investigate the N-S asymmetries of the filament numbers, areas, spine lengths, and tilt angles. The N-S asymmetry of filament features has been calculated by the formulas
\begin{equation}\label{eq4}
A_{N}=\frac{N_{N}-S_{N}}{N_{N}+S_{N}},
 \end{equation}

\begin{equation}\label{eq5}
A_{C}=\frac{N_{C}-S_{C}}{N_{C}+S_{C}},
 \end{equation}
which have been employed by many authors \citep{1996Duchlev,2001Duchlev,2006Zharkov,2009Li1,2009Joshi,2010Bankoti,2011Bankoti}. $A_{N}$ is the normalized N-S asymmetry index, $N_{N}$  is the number of the filament features in the northern hemisphere and $S_{N}$ is that in the southern hemisphere.  $A_{C}$ is the normalized cumulative N-S asymmetry index, $N_{C}$  is the cumulative value of the filament features in the northern hemisphere and $S_{C}$ is that in the southern hemisphere. If $A_{N}>0$ (or $A_{C}>0$), the northern hemisphere is the dominant hemisphere and for $A_{N}<0$ (or $A_{C}<0$), the reverse is true. We calculated the yearly asymmetries of the filament numbers in different latitude bands from 1988 to 2013, then plotted in Figure~\ref{fig17}. All filament numbers in the period from 1989 to 1992 show that the southern hemisphere is the dominant hemisphere. From 1993 to 2005 the northern hemisphere is the dominant hemisphere. However, the asymmetries of these years being relatively small indicate that there is no obvious difference between the northern and southern hemispheres in these years. From 2006 to 2008 the southern hemisphere is the dominant one, which is opposite to that from 2009 to 2011, after that it changes back to the southern hemisphere in 2012 and 2013. The asymmetry indices  $A_{N}$ for the years from 2005 to 2013  have relatively large variations showing that the northern and southern hemispheres have relatively large differences in filament numbers. The asymmetries of the filament numbers in the latitude bands $0^{\circ}  \sim  50^{\circ}$ similar to that in all latitudes due to the most of the filament are detected in low latitudes. While the asymmetries of the filament numbers in high latitude bands have relatively dramatic changes. Besides, in order to keep consistent with the previous grouping method, we divide the filaments into three groups based on the periods within the three solar cycles. Table~\ref{table7} lists the asymmetry of the detected filament numbers in the latitude bands $0^{\circ}  \sim  50^{\circ}$ and $50^{\circ}  \sim  90^{\circ}$, and all latitudes from 1988 to 2013 within the three solar cycles.  It is found that the N-S asymmetry indices $A_{N}$ are all negative in solar cycle 22, which indicates that the dominant hemisphere is the southern hemisphere. It becomes opposite in solar cycle 23, namely, the northern hemisphere is the dominant one. For the period in the rising phase of solar cycle 24, we can see that for the filament numbers in the latitude bands $50^{\circ}  \sim  90^{\circ}$ the dominant hemisphere is the southern hemisphere, whereas the filament numbers in the latitude band $0^{\circ}  \sim  50^{\circ}$ and in all latitudes show the opposite results.

The area range is divided into two parts by the area size $5.0\times10^{8}$ km$^{2}$. We then calculate the yearly N-S asymmetry indices $A_{N}$ and $A_{C}$  in the two area ranges in low, high and all latitude bands. The results are plotted in Figures~\ref{fig18}--\ref{fig19}. From Figure~\ref{fig18} we find that in the periods from 1988 to 1996 and from 2005 to 2013 the N-S asymmetry indices $A_{N}$ have relatively large variations in the latitude bands $0^{\circ}  \sim  50^{\circ}$ and all latitudes. While that in the period from 1997 to 2004 have relatively small variations. It indicates that the differences between the northern and southern hemispheres in solar cycle 22 and the rising phase of solar cycle 24 are bigger than that in solar cycle 23. The asymmetries of the filament numbers in high latitude bands have relatively dramatic changes, especially for the filaments with areas exceeding $5.0\times10^{8}$ km$^{2}$. Noticed that the N-S asymmetry indices $A_{N}$ of the filaments with areas exceeding $5.0\times10^{8}$ km$^{2}$ are equal to 1 (-1) that is due to the less filaments in high latitudes. The N-S asymmetry indices $A_{C}$ in Figure~\ref{fig19} show the similar results with  $A_{N}$. Tables~\ref{table8} and \ref{table9} list the specific results of the N-S asymmetry of the filament numbers and the cumulative filament areas within the two area ranges in the latitude bands $0^{\circ}  \sim  50^{\circ}$ and $50^{\circ}  \sim  90^{\circ}$ within the three solar cycles, respectively. In order to give an intuitionistic view of the N-S asymmetry index variations of the filament numbers and the cumulative filament areas, we also plot the results in low and high latitude bands in Figure~\ref{fig20}.  Figure~\ref{fig20} and Tables~\ref{table9}--\ref{table10} indicate that the N-S asymmetry of the filament numbers and the cumulative filament areas have similar trends. The dominant hemisphere is the southern hemisphere in solar cycle 22 and it becomes to be the northern hemisphere in solar cycle 23 and the rising phase of solar cycle 24 in low latitude bands.  In high latitude bands the dominant hemisphere is the northern hemisphere in solar cycles 22 and 23 and it becomes to be the southern hemisphere in the rising phase of solar cycle 24. From Figure~\ref{fig20}(a, c) we also find that the differences between the northern and southern hemispheres in solar cycle 23 are less than that in solar cycle 22 and the rising phase of the solar cycle 24. In solar cycle 22 and the rising phase of solar cycle 24, the results show that the dominant hemisphere is opposite between the low and high latitudes. While for solar cycle 23 the northern hemisphere is dominant in all latitudes. 

The filament spine length is arranged into the short and long range by $5.0\times10^{4}$ km. The yearly N-S asymmetry indices $A_{N}$ and $A_{C}$  in the two spine length ranges in low, high and all latitude bands  are plotted in Figures~\ref{fig21}--\ref{fig22}. The N-S asymmetry of the number of the filaments with spine lengths and the cumulative filament spine lengths are calculated and demonstrated in Tables~\ref{table10} and \ref{table11}. The two tables list the specific results of the N-S asymmetry of the filament numbers and the cumulative spine lengths in the latitude bands $0^{\circ}  \sim  50^{\circ}$ and $50^{\circ}  \sim  90^{\circ}$ from 1988 to 2013 within the three solar cycles, respectively. The N-S asymmetry index variations within two spine length ranges  in the latitude bands $0^{\circ}  \sim  50^{\circ}$ and $50^{\circ}  \sim  90^{\circ}$ are plotted in Figure~\ref{fig23}. The results is similar to that of the filament areas. It is shown that the dominant hemisphere is the southern hemisphere in solar cycle 22 and becomes the northern hemisphere in solar cycle 23 and the rising phase of solar cycle 24 in low latitudes. In high latitudes the dominant hemisphere in solar cycle 22 and the rising phase of solar cycle 24 is opposite to that in low latitude bands. The northern hemisphere is the dominant hemisphere in solar cycle 23 in all latitudes. But the variation does not change drastically, it means that the difference between the two hemispheres is not significant.

Figure~\ref{fig24} shows the yearly N-S asymmetry indices $A_{N}$ in different tilt angles. The filament tilt angle definition is the same as in Section~\ref{tiltangle}. We can see that most indices  $A_{N}$ is positive (negative) when the tilt angles is negative (positive) in low latitude bands. It indicated that most filaments with negative (positive) tilt angles dominant the northern (southern) hemisphere in low latitude bands. However, the variation changes drastically in high latitudes which has no obvious rule. Table~\ref{table12} lists the N-S asymmetry indices $A_{N}$ variations with the filament tilt angles.  These tables give the specific results of the N-S asymmetry of the filament numbers within different tilt angle ranges in the latitude bands $0^{\circ}  \sim  50^{\circ}$ and $50^{\circ}  \sim  90^{\circ}$ from 1988 to 2013 in the three solar cycles, respectively.  Figure~\ref{fig25} shows the N-S asymmetry index variations within different tilt angle ranges in the low and high latitude bands. The variation profiles are similar to sine functions, which indicates that the dominant orientation of the northern (southern) hemisphere is $-90^{\circ}  \sim  0^{\circ}$ ($0^{\circ}  \sim  90^{\circ}$). On the contrary, the N-S asymmetry index variations profiles in high latitudes are similar to cosine functions, though some parts are not very smooth, such as that in the rising phase of solar cycle 24. It means that in high latitude bands the dominant orientation of the northern (southern) hemisphere is $0^{\circ}  \sim  90^{\circ}$ ($-90^{\circ}  \sim  0^{\circ}$), which is opposite to that in low latitude bands.

\section{Discussion and Conclusion}
\label{conclusion}

\begin{figure}
\centering
\includegraphics[width=1.0\textwidth,clip=]{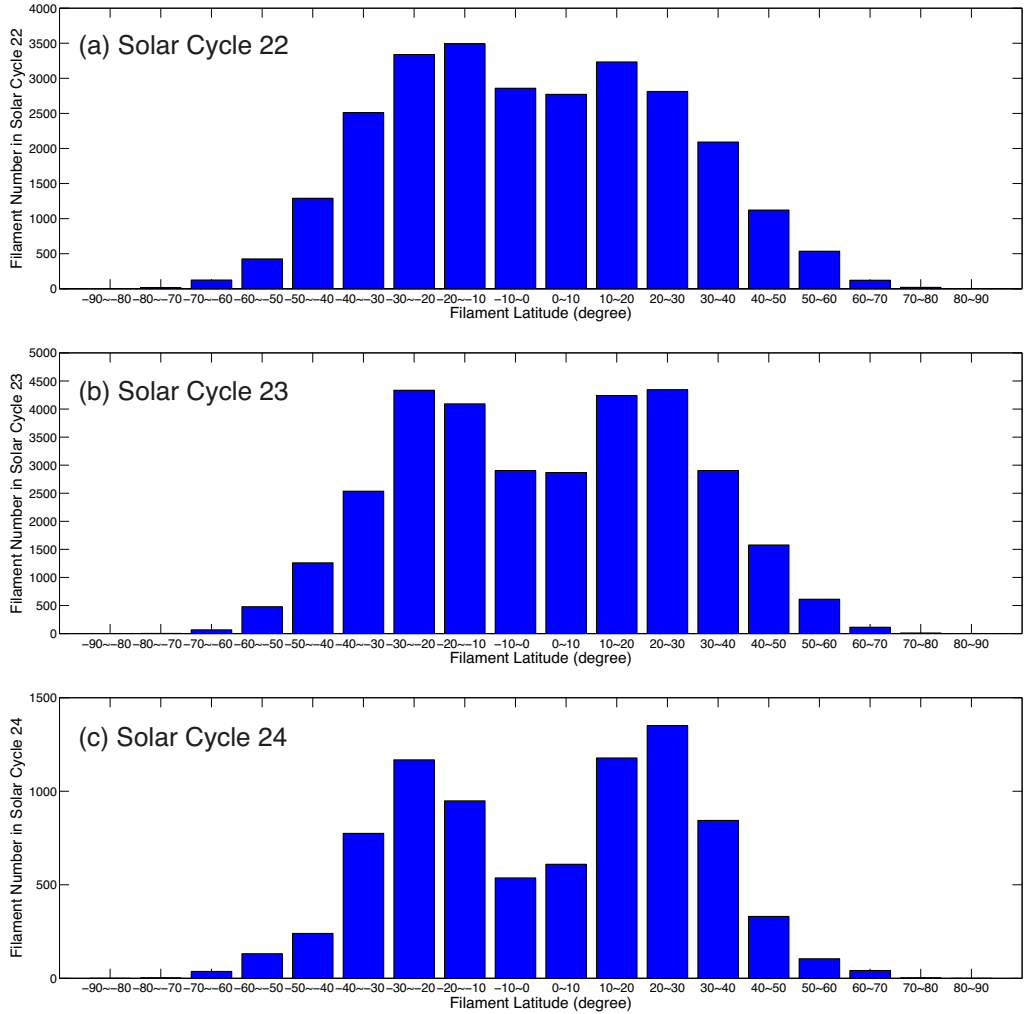}
\caption{Distributions of the filament numbers in different latitude bands. Panels (a), (b), and (c) show the distribution of the filament numbers in the different latitude bands in solar cycles 22, 23, and the rising phase of solar cycle 24, respectively.
}
\label{fig1}
\end{figure}

\begin{figure}
\centering
\includegraphics[width=1.0\textwidth,clip=]{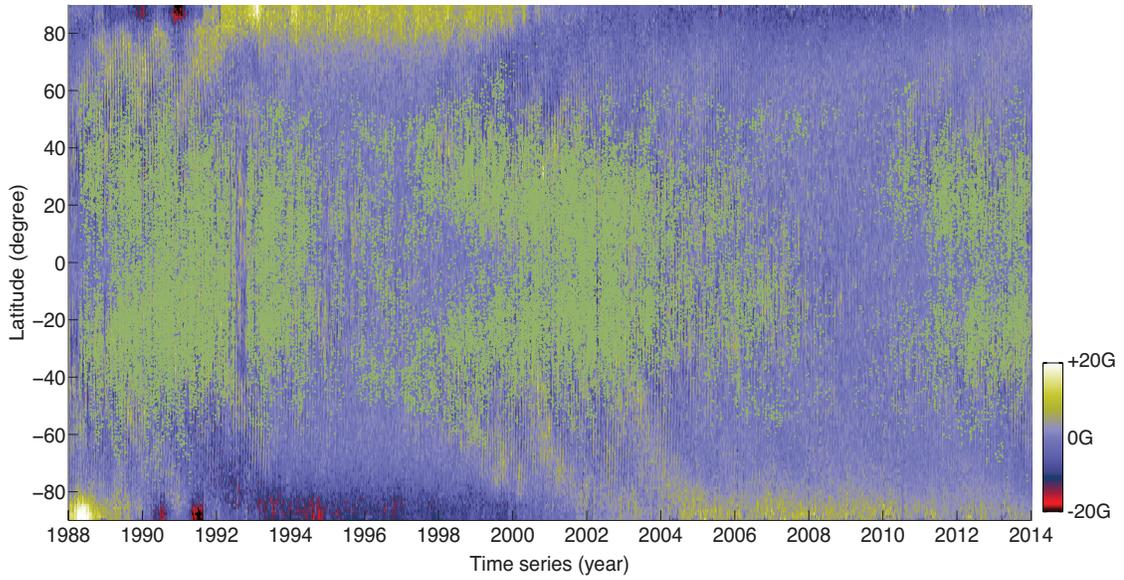}
\caption{Butterfly diagram of filaments from April 1988 to December 2013. Each green dot represents a single observation. The background is the butterfly diagram of magnetic fields during the same time periods for comparison.  It illustrates Hale's Polarity Law visibly.
}
\label{fig2}
\end{figure}

\begin{figure}
\centering
\includegraphics[width=1.0\textwidth,clip=]{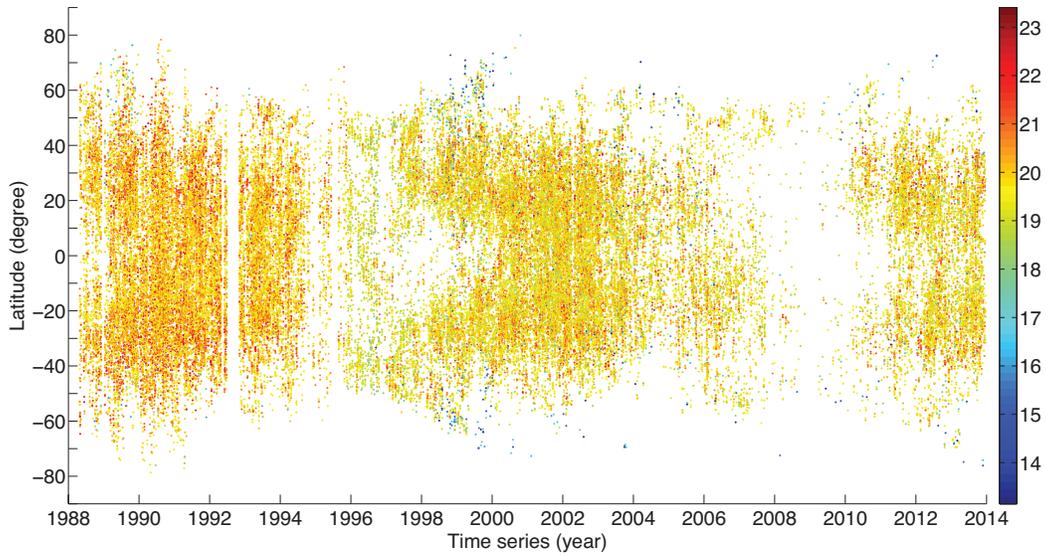}
\caption{Similar to Figure~\ref{fig2} but the colors show the decadic logarithm of filament areas.
}
\label{fig3}
\end{figure}

We adopt our improved automated filament detection method to process 6564 images which are mainly obtained by BBSO from April 1988 to December 2013. As a result, $67,432$ filaments are detected in total. We divide the filaments into 3 groups, according to solar cycles 22 to 24. The detected numbers of the filaments are 26392, 32633, and 8418 in these three periods, respectively. The temporal evolutions of the latitudinal distributions of these filaments are depicted as the scatter plots in Figures~\ref{fig1}--\ref{fig3}, where we can clearly see the butterfly diagrams. The butterfly diagram of the filaments is compared with that of the magnetic fields, and an agreement is seen between them, though the butterfly diagram of the filament is wider in latitude than that of the magnetic field. This is not surprising since filaments represent the PILs that divide the magnetic fields of opposite polarities. Besides the butterfly diagram, we also see the poleward migration of high-latitude filaments in association with the poleward migration of the magnetic field. A similar work was done by \citet{1983Makarov}, who demonstrated that the evolution and migration of the magnetic field can be investigated by monitoring the migration of the PILs based on the H$\alpha$ synoptic charts. The filament number in solar cycle 22 is less than that in solar cycle 23, which is opposite to the result that the sunspot activity in solar cycle 23 is weaker than that in solar cycle 22. It is because that solar cycle 22 has 721 images less than in solar cycle 23. Besides the filament number and location, we also analyzed the area, spine length, tilt angle, and barb number of the filaments.

The filament area is divided into four ranges. It is found that over $80\%$ of all the filaments have an area less than $1.0\times10^{9}$ km$^{2}$. The filaments with areas less than $2.5\times10^{8}$ km$^{2}$ almost dominate the period from 1996 to 1998 and those with areas of $2.5\times10^{8} \sim 5.0\times10^{8}$ km$^{2}$ nearly dominate in the period from 1999 to 2007. It indicates that from the beginning to the maximum of solar cycle 23 the filaments are relatively smaller and from the maximum to the minimum the filaments become relatively larger. Notice that the filaments with areas less than  $1.0\times10^{9}$ km$^{2}$ dominate in the latitude band $0^{\circ}  \sim 60^{\circ}$ and the filaments with areas less than  $5.0\times10^{8}$ km$^{2}$ dominate in the latitude band $60^{\circ}  \sim 90^{\circ}$. It indicates that the filaments in low latitudes have diverse areas but the filaments in high latitudes are relative smaller. There are two possibilities:  one is that when large filaments migrate poleward, they may break into several pieces and the filament fragments are far away from each other, our program cannot figure out if they belong to a single filament; the other is that there are many really small filaments appearing in high latitudes. 

The filament spine length is divided into three groups. It is found that over $80\%$ of all detected filaments have lengths less than $1.0\times10^{5}$ km. For each group, the latitudinal distribution of the filaments is also bimodal. The filament number peaks nearly appear in the latitude band  $10^{\circ} \sim 30^{\circ}$ in both hemispheres throughout the three solar cycles. We also find that the shorter filaments mainly appear in high latitudes, which is similar to the results of the filament area. It is not difficult to understand since the filament spine length and area are both associated with the filament size.

From the results for the tilt angle directions in both hemispheres, we found that the northeast direction is the dominant orientation in the northern hemisphere and the southeast direction is the dominant orientation in the southern hemisphere in all three solar cycles. It has been known that filaments with a right-handed twist magnetic flux rope, namely sinistral chirality, mainly appear in the southern hemisphere and those with left-handed twist (dextral chirality) in the northern hemisphere \citep{1998Martin}. Filaments with dextral chirality in the northern hemisphere are usually oriented in the northeast direction and those with sinistral chirality in the southern hemisphere are usually oriented in the southeast direction. Therefore, our results indirectly confirm the hemispheric rule, i.e., most filaments in the northern hemisphere have dextral chirality with negative helicity and most filaments in the southern hemisphere have sinistral chirality with positive helicity. It is also found that the filaments within the latitude band $0^{\circ} \sim 50^{\circ}$ are mostly oriented in the northeast direction in the northern hemisphere and southeast direction in the southern hemisphere, whereas the northwest is the dominant direction in the northern hemisphere and the southwest is the dominant direction in the southern hemisphere for the filaments within the latitude band $50^{\circ} \sim 90^{\circ}$ throughout the three solar cycles. However, the variations in high latitudes do not as significant as that in low latitudes. These results indicate that the solar magnetic fields may have complex and different evolving processes in the low and high latitudes. We calculated the specific tilt angle distributions in different latitude bands. About $80\%$ of all the filaments have tilt angles within $[0^{\circ}, 60^{\circ}]$ in both hemispheres. Some authors analyzed the tilt angles of the bipolar magnetic regions \citep{1919Hale,2012Stenflo}, where the tilt angle is defined as the angle between the equator and a line connecting the geometric centers of the conjugate polaritie. It was found that emerging bipoles favor the northeast-southwest orientation in the northern hemisphere and the southwest-northwest orientation in the southern hemisphere, i.e., Hale's law. Since the filaments are always aligned with photospheric magnetic PILs \citep{1998Martin}, the filament spine would be approximately perpendicular to the line connecting the two polarities of the bipole and deviates from Hale's law if the filament is formed between the preceding and following polarities of a bipole. Our result that filaments have the same preference of orientation as sunspot pairs implies that most filaments are formed either at the boundary of an active region or at the interface between two active regions. Besides, the bipolar tilt angle grows as the latitude increases, which is known as Joy's law \citep{1919Hale}. In our study, the filament tilt angle does not change significantly with latitude from the year 1988 to 2013. The difference from Joy's law can again be understood to be due to the fact that most filaments are not formed along the main PILs inside bipolars. Only some active region filaments are along the PILs between the preceding and following popalities, whereas most filaments, especially those quiescent filaments, are generally located above the PILs in decayed active regions or between two active regions.

The latitudinal distribution of the filaments with different numbers of barbs is also biomodal, and the peaks appear around  $10^{\circ} \sim 30^{\circ}$ in both hemispheres, which is similar to the results for the filament area and spine length. About $75\%$ of all the filaments have less than 5 barbs.

We calculated the monthly mean latitudes of the filaments with latitude in the range of $0^{\circ} \sim 50^{\circ}$ and $>50^{\circ}$ in both hemispheres. Then we adopted the cubic polynomial fitting to get the drift velocities of the filaments. The latitudinal migrations of the filaments experience three stages in both solar cycles 22 and 23: from the beginning to the maximum of each solar cycle, the drift velocity is relatively higher. From the solar maximum to the years before the solar minimum the drift velocity becomes relatively lower. After that the migration becomes divergent and more filaments appear in high latitudes, marking the start of a new cycle. The calculated drift velocities in solar cycle 23 are similar to those in our previous work \citep{2013Hao}. These results are similar to the statistical results of \citet{2010Li}. The filaments with latitudes lower than $50^{\circ}$ migrate toward the equator in both hemispheres during the whole solar cycles. Since most of the detected filaments, i.e., over 90\%, are located lower than $50^{\circ}$, filaments migrate toward the equator on average. It is noticed that most filaments with latitudes higher than $50^{\circ}$ migrate toward the polar regions with relatively high velocities, namely the so-called ``rush to the poles"\citep{1982Topka,1989Makarov,1997Altrock,2006Shimojo,2008Li,2010Li}, especially during the time intervals from 1988 to 1989 and from 1997 to 2000, i.e., from the beginning to the year around the maximum. Then they change the drift direction toward the equator after the maximum. We also calculated the drift velocities of the filaments in the latitudes around $50^{\circ}$, these filaments may migrate toward either the equator or the polar region. These results indicate that the filaments in the medium latitudes around $50^{\circ}$ can migrate toward both directions and some filaments in high latitudes can also migrate toward the equator. It is noticed that both the drift velocity and the dispersion of the filaments in the rising phase of the solar cycle decrease from solar cycles 22 to 23. Since the length of solar cycle 23 is longer than solar cycle 22, we surmise that the drift velocity may be correlated with the length of the solar cycle: the shorter the solar cycle is, the higher the filament drift velocity is. Interestingly, \citet{2003Hathaway} compared the drift rate of sunspots at the maximum of sunspot cycle with the period of each cycle for each hemisphere and found that hemispheres with faster drift rates have shorter periods. These observational features imply that a deep meridional flow toward the equator is the primary driver of the sunspot cycle, which influences the amplitude of the following cycle\citep{2003Hathaway}. After the solar maximum, the drift velocities become lower, with an amplitude of $0 \sim 1.0$ m s$^{-1}$. Besides, in order to find the relationship between the drift velocity and the filament area and the spine length, we also calculated the drift velocities of the filaments within different area ranges and spine length ranges, respectively. However, the results show that the drift velocity of the filaments changes slightly with the increasing area or spine length. In other words, the filament drift velocity is nearly independent of the filament area or the spine length.

\begin{figure}
\centering
\includegraphics[width=1.0\textwidth,clip=]{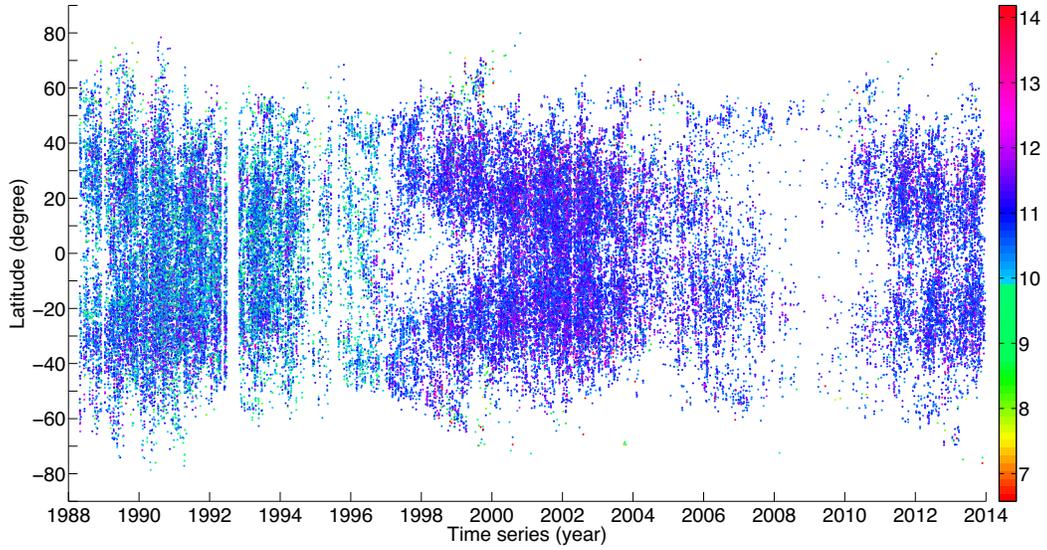}
\caption{Similar to Figure~\ref{fig2} but the colors show the decadic logarithm of filament spine lengths.
}
\label{fig4}
\end{figure}

\begin{figure}
\centering
\includegraphics[width=1.0\textwidth,clip=]{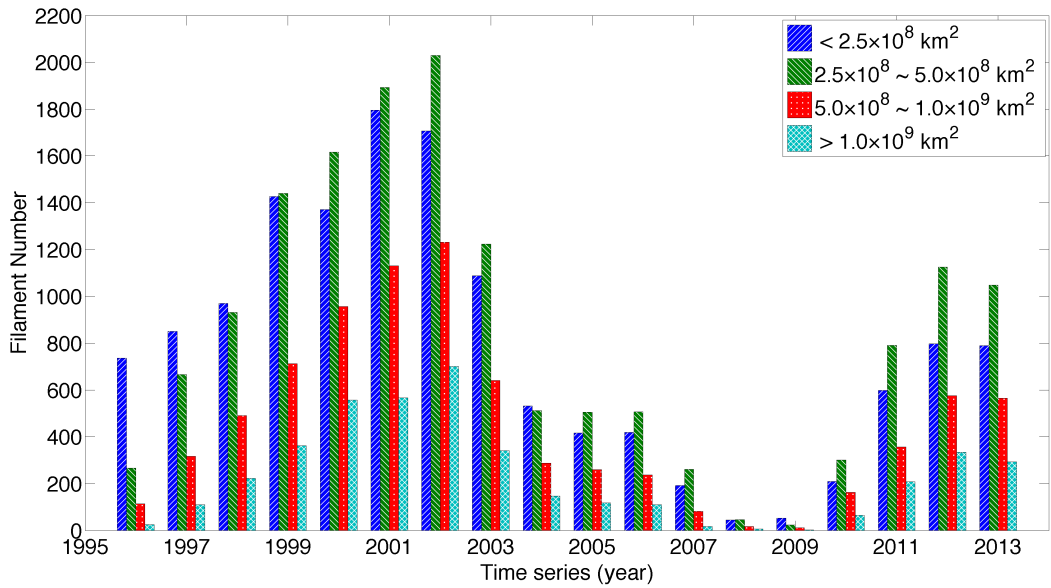}
\caption{Distributions of the filament areas from 1996 to 2013.
}
\label{fig5}
\end{figure}

\begin{figure}
\centering
\includegraphics[width=1.0\textwidth,clip=]{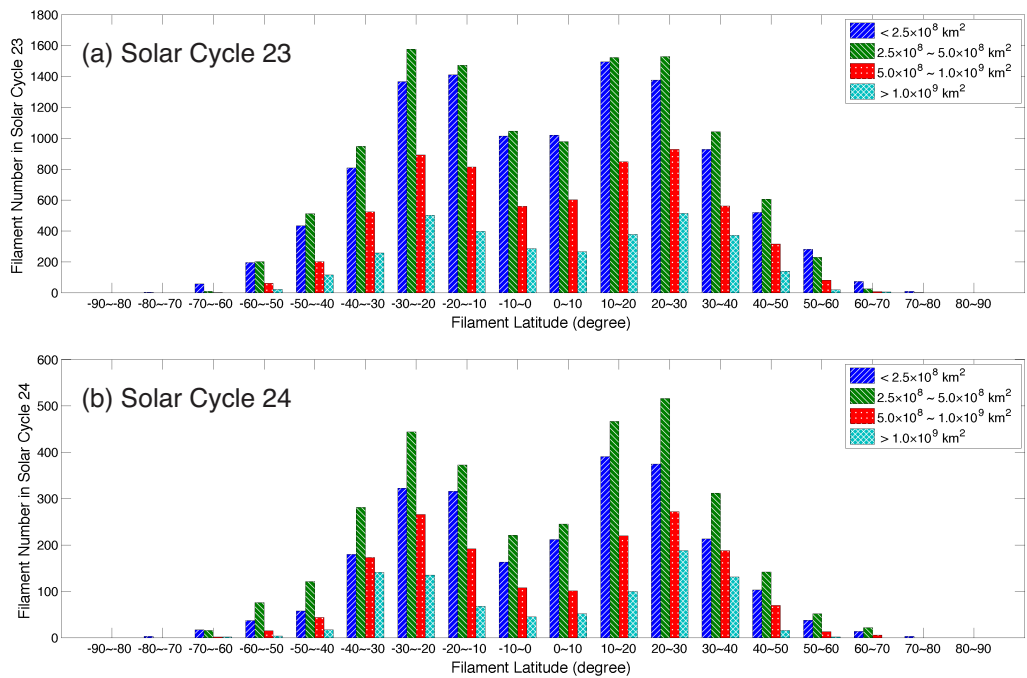}
\caption{Distributions of the filament areas in the different latitude bands. Panels (a) and (b)show the distributions of the filament areas in different latitude bands in solar cycle 23 and the rising phase of solar cycle 24, respectively.
}
\label{fig6}
\end{figure}

\begin{figure}
\centering
\includegraphics[width=1.0\textwidth,clip=]{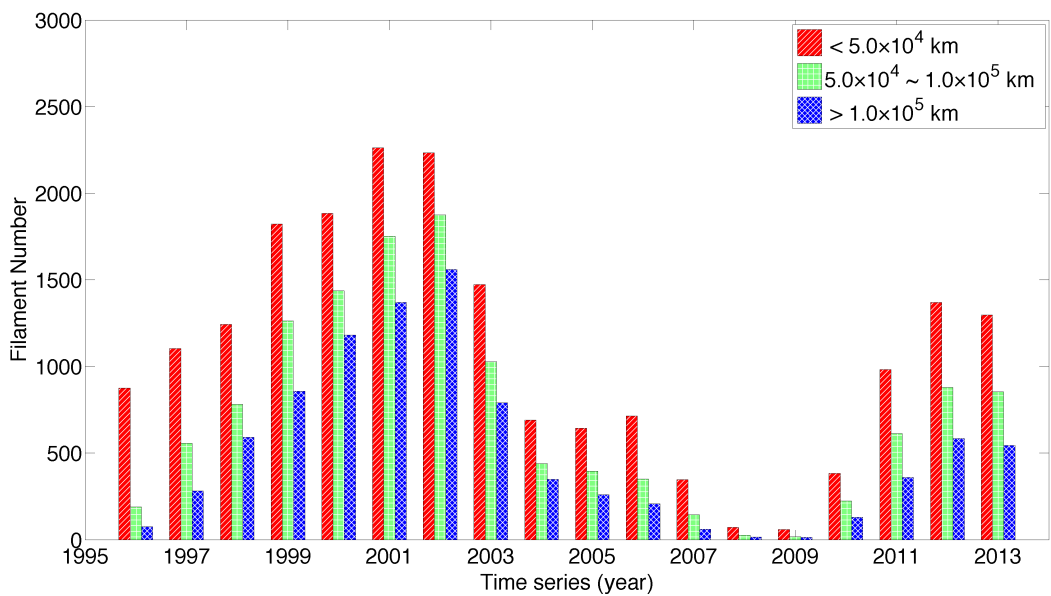}
\caption{Distributions of the filament spine lengths from 1996 to 2013.
}
\label{fig7}
\end{figure}

\begin{figure}
\centering
\includegraphics[width=1.0\textwidth,clip=]{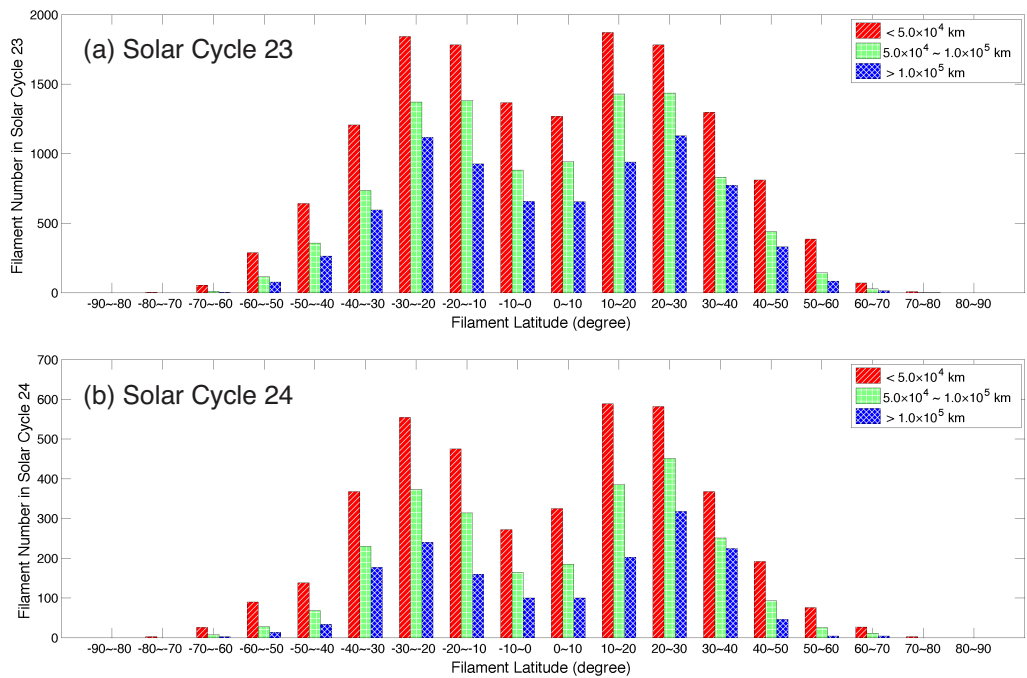}
\caption{Distributions of the filament spine lengths in different latitude bands. Panels (a) and (b) show the distributions of the filament lengths in different latitude bands in solar cycle 23 and the rising phase of solar cycle 24, respectively.
}
\label{fig8}
\end{figure}

\begin{figure}
\centering
\includegraphics[width=1.0\textwidth,clip=]{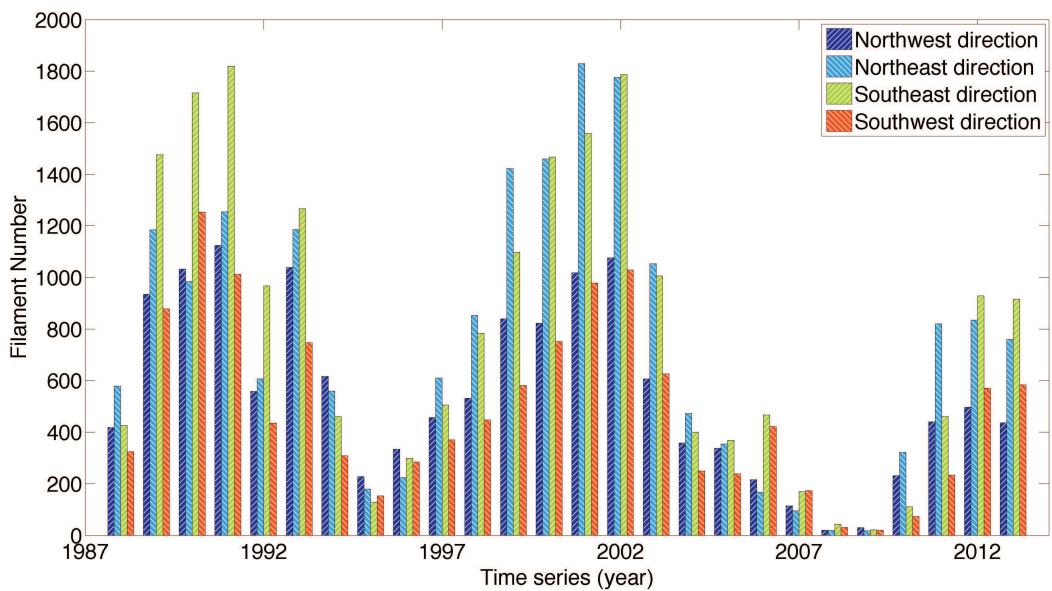}
\caption{Distributions of the filament tilt angle directions from 1988 to 2013.
}
\label{fig9}
\end{figure}

\begin{figure}
\centering
\includegraphics[width=1.0\textwidth,clip=]{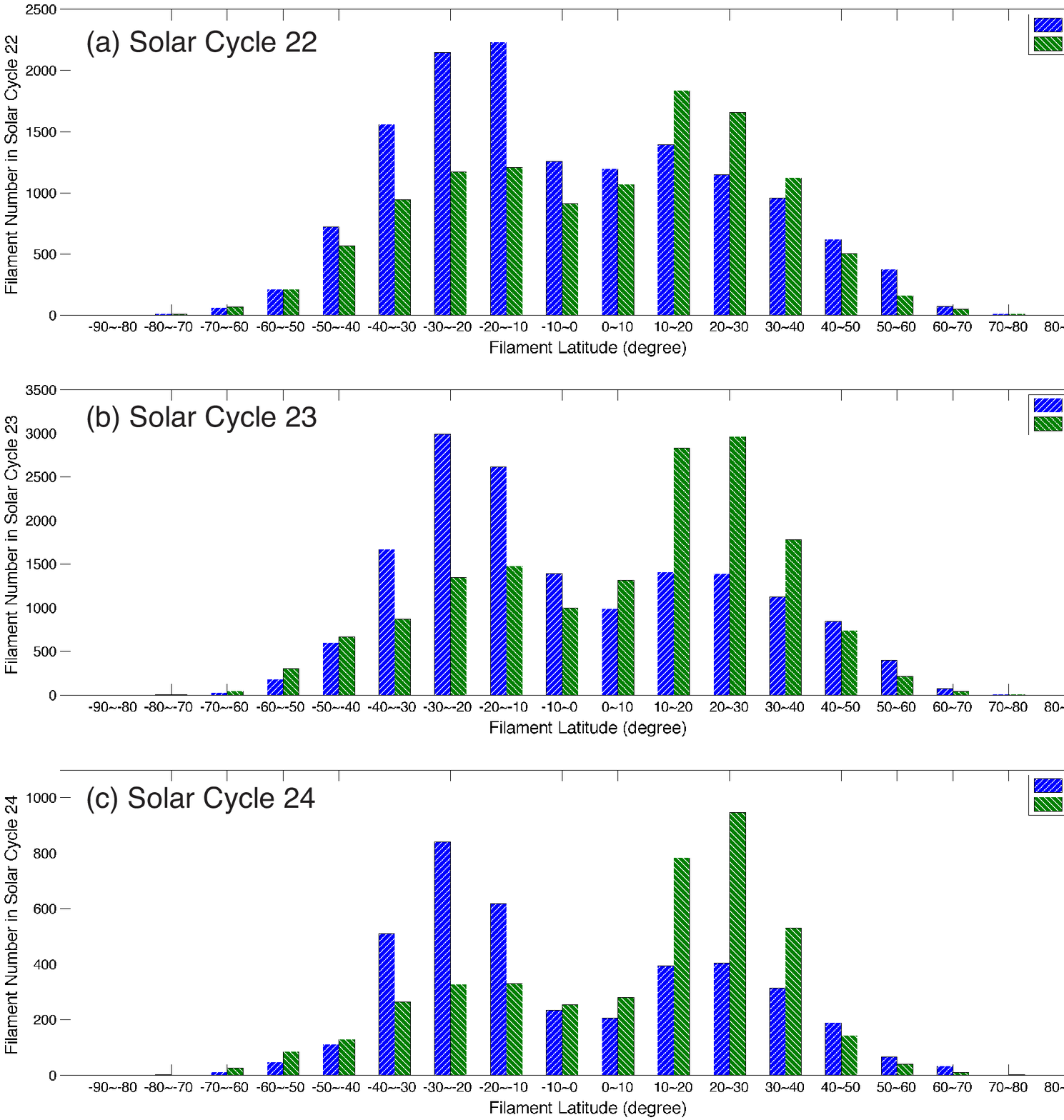}
\caption{Distributions of the filament tilt angle direction in different latitude bands. Panels (a), (b) and (c) show the distributions of the filament tilt angle direction in different latitude bands in solar cycles 22, 23, and the rising phase of solar cycle 24, respectively.
}
\label{fig10}
\end{figure}

\begin{figure}
\centering
\includegraphics[width=1.0\textwidth,clip=]{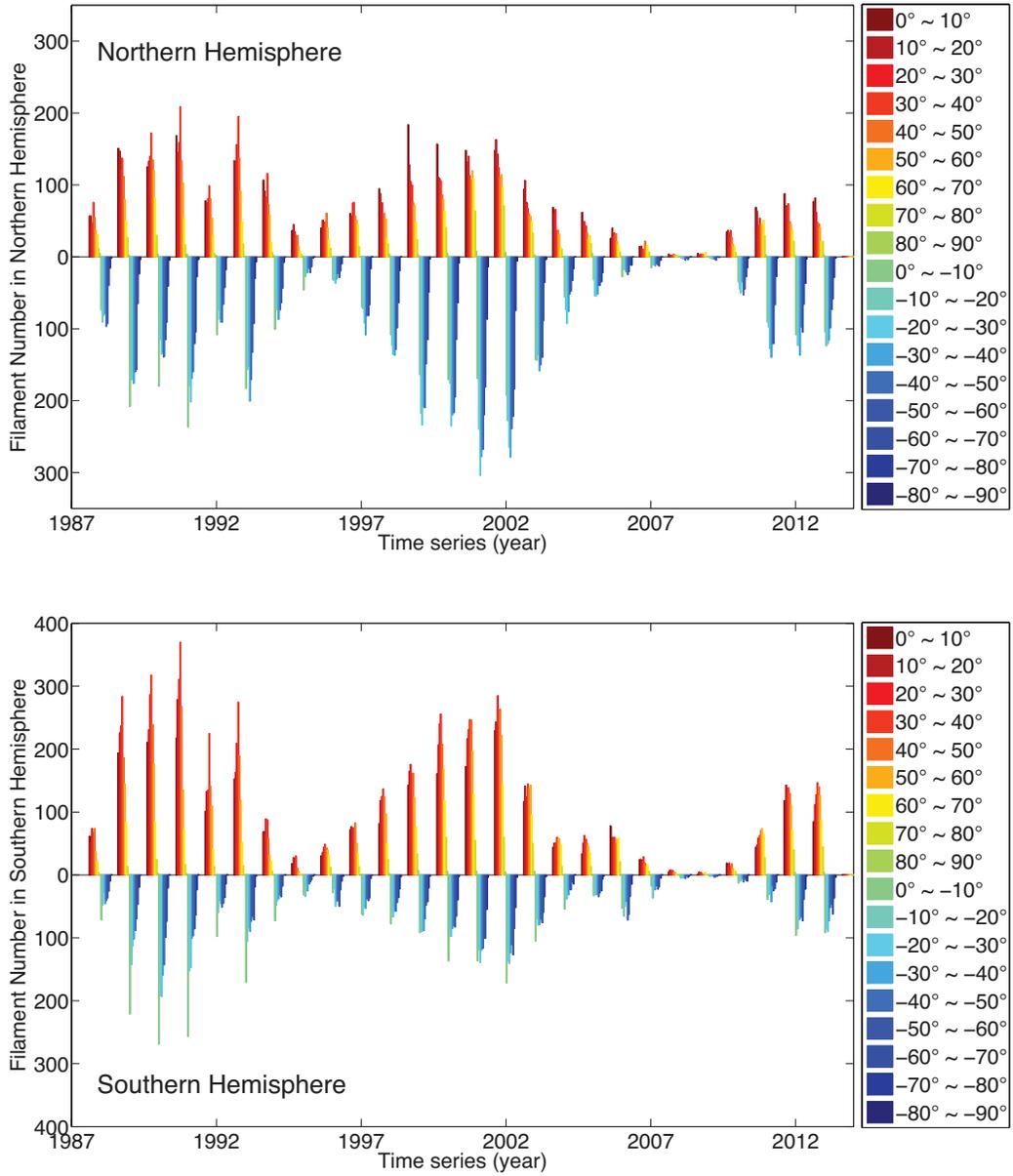}
\caption{Distributions of filament tilt angles in the northern and southern hemispheres from 1988 to 2013.
}
\label{fig11}
\end{figure}

\begin{figure}
\centering
\includegraphics[width=1.0\textwidth,clip=]{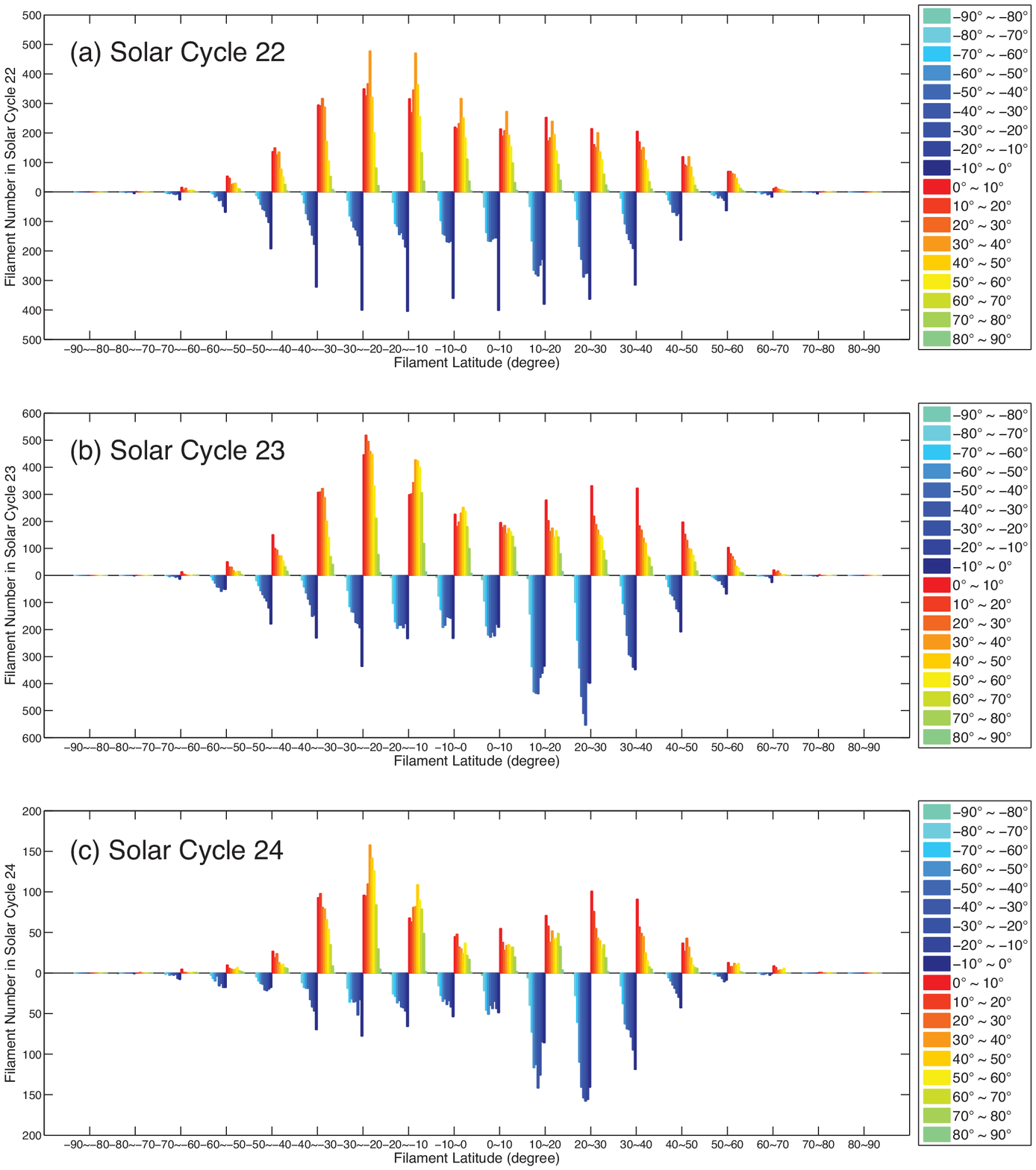}
\caption{Distributions of the filament tilt angles in different latitude bands. Panels (a), (b) and (c) show the distribution of the filament tilt angles in different latitude bands in solar cycles 22, 23, and the rising phase of solar cycle 24, respectively.
}
\label{fig12}
\end{figure}

\begin{figure}
\centering
\includegraphics[width=1.0\textwidth,clip=]{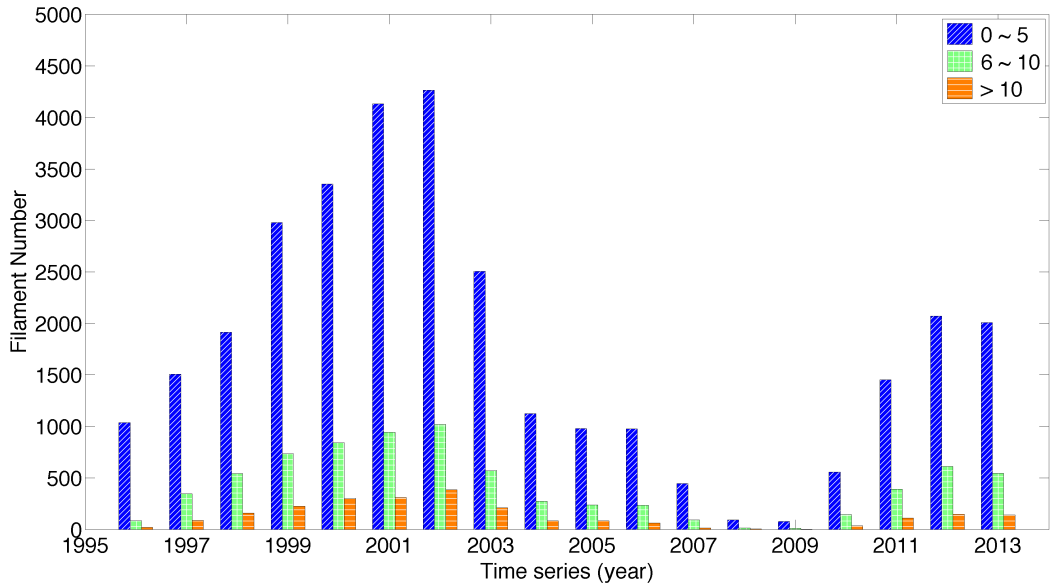}
\caption{Distributions of filament barb numbers from 1996 to 2013.
}
\label{fig13}
\end{figure}

\begin{figure}
\centering
\includegraphics[width=1.0\textwidth,clip=]{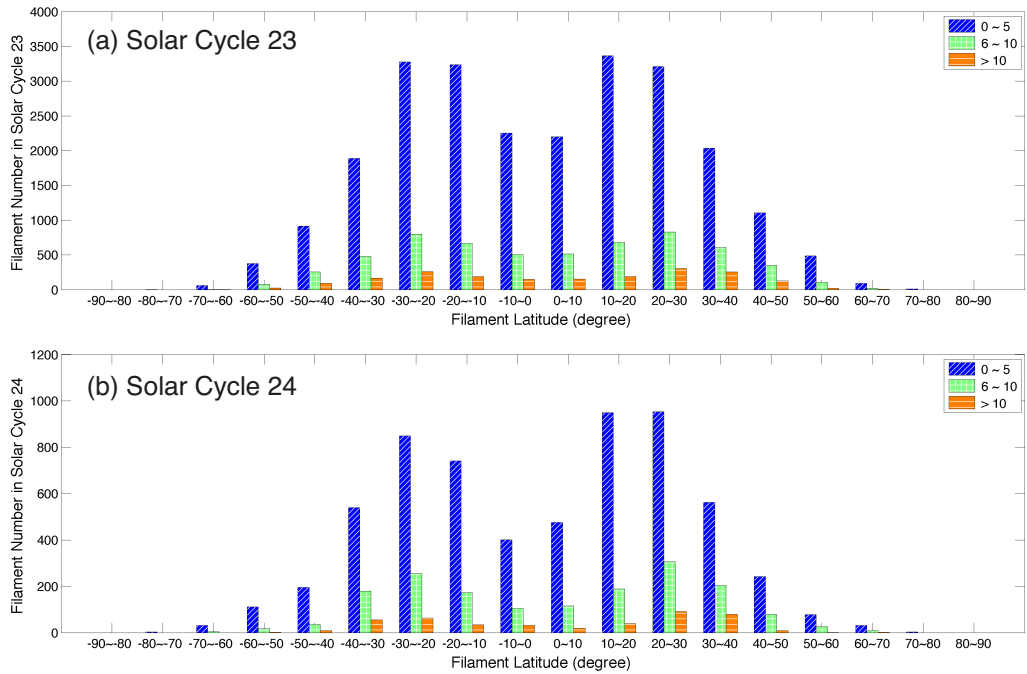}
\caption{Distributions of filament barb numbers in different latitude bands. Panels (a) and (b) show the distributions of the filament barb numbers in different latitude bands in solar cycle 23 and the rising phase of solar cycle 24, respectively.
}
\label{fig14}
\end{figure}

\begin{figure}
\centering
\includegraphics[width=1.0\textwidth,clip=]{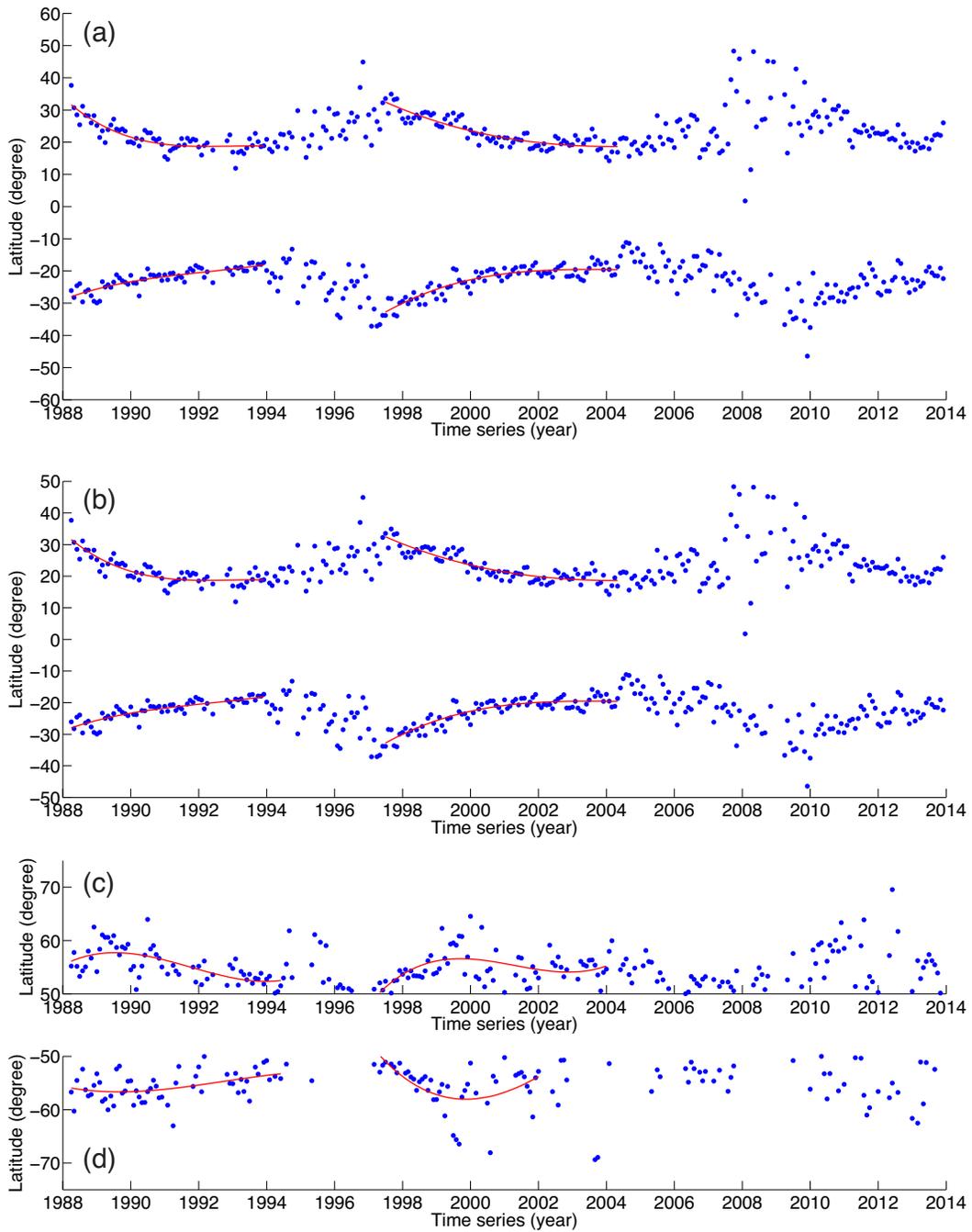}
\caption{Temporal evolution of the monthly mean latitude distributions of the filaments from April 1998 to December 2013. Panels (a), (b), (c), and (d) show the monthly mean latitude distributions of the filaments in all latitudes, the latitude bands $0^{\circ} \sim 50^{\circ}$, $50^{\circ} \sim 90^{\circ}$ in the northern hemisphere and $50^{\circ} \sim 90^{\circ}$ in the southern hemisphere, respectively. Cubic polynomial fittings are shown as red lines.
}
\label{fig15}
\end{figure}

\begin{figure}
\centering
\includegraphics[width=1.0\textwidth,clip=]{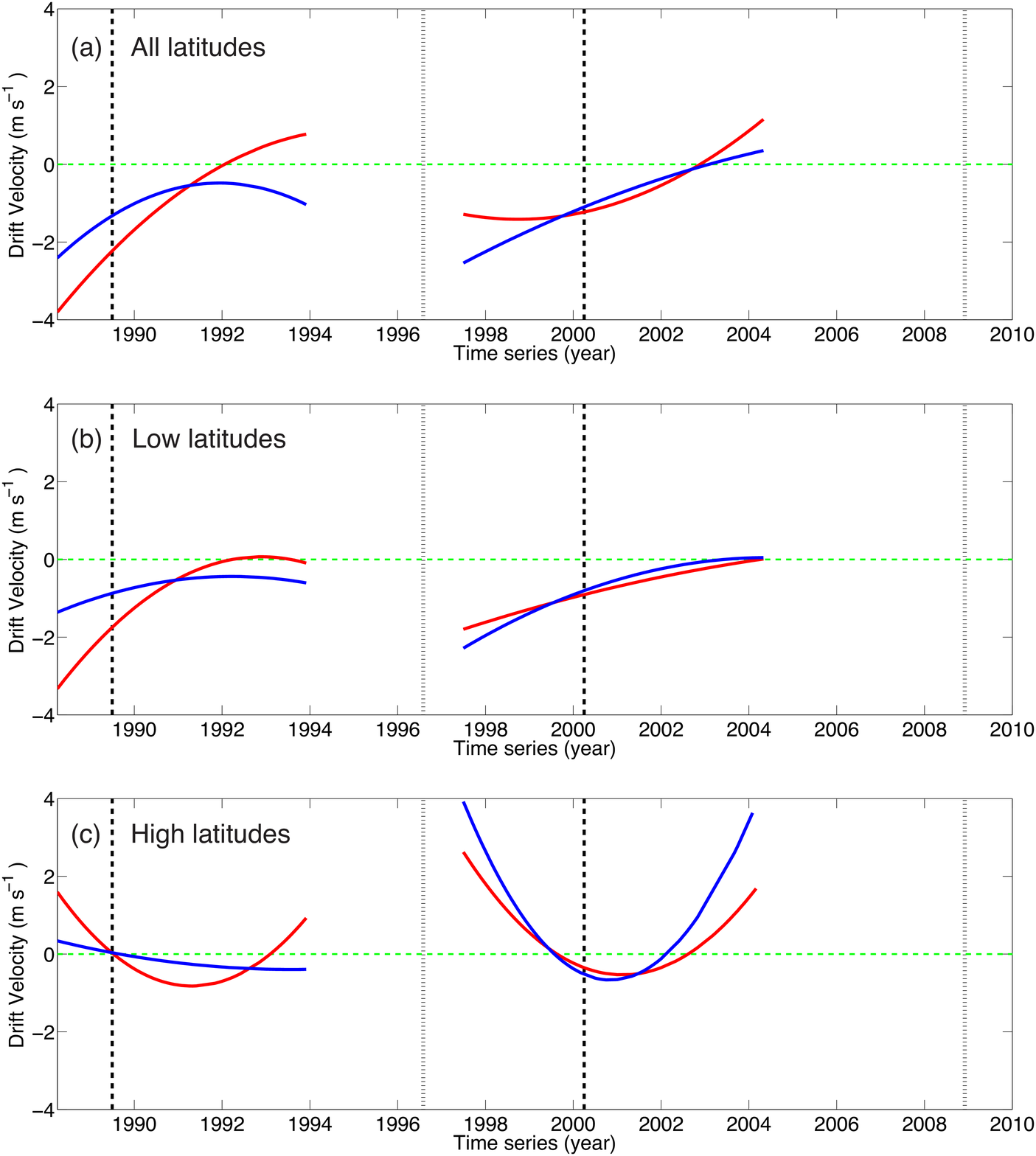}
\caption{Drift velocity variations of the monthly mean latitude distributions of the filaments in solar cycle 22 and 23. Panels (a), (b), and (c) show the drift velocity variations in all latitudes, the latitude bands $0^{\circ} \sim 50^{\circ}$ and $50^{\circ} \sim 90^{\circ}$. The red lines and the blue lines indicate the drift velocities of the filaments in the northern and southern hemispheres, respectively. The black dashed lines and the black dotted line show the solar maximum and solar minimum, respectively. The green dashed lines are the zero velocity lines for reference.
}
\label{fig16}
\end{figure}

\begin{figure}
\centering
\includegraphics[width=1.0\textwidth,clip=]{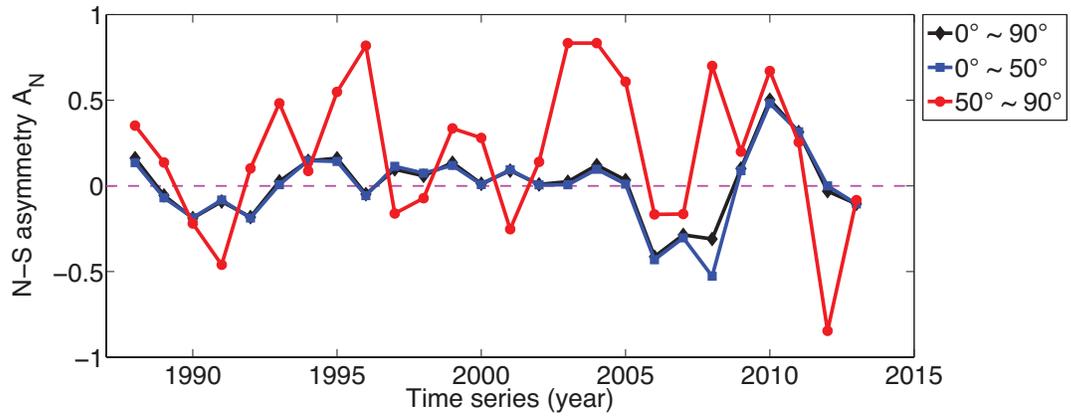}
\caption{The yearly asymmetries of the filament numbers in different latitude bands from 1988 to 2013.
}
\label{fig17}
\end{figure}

\begin{figure}
\centering
\includegraphics[width=1.0\textwidth,clip=]{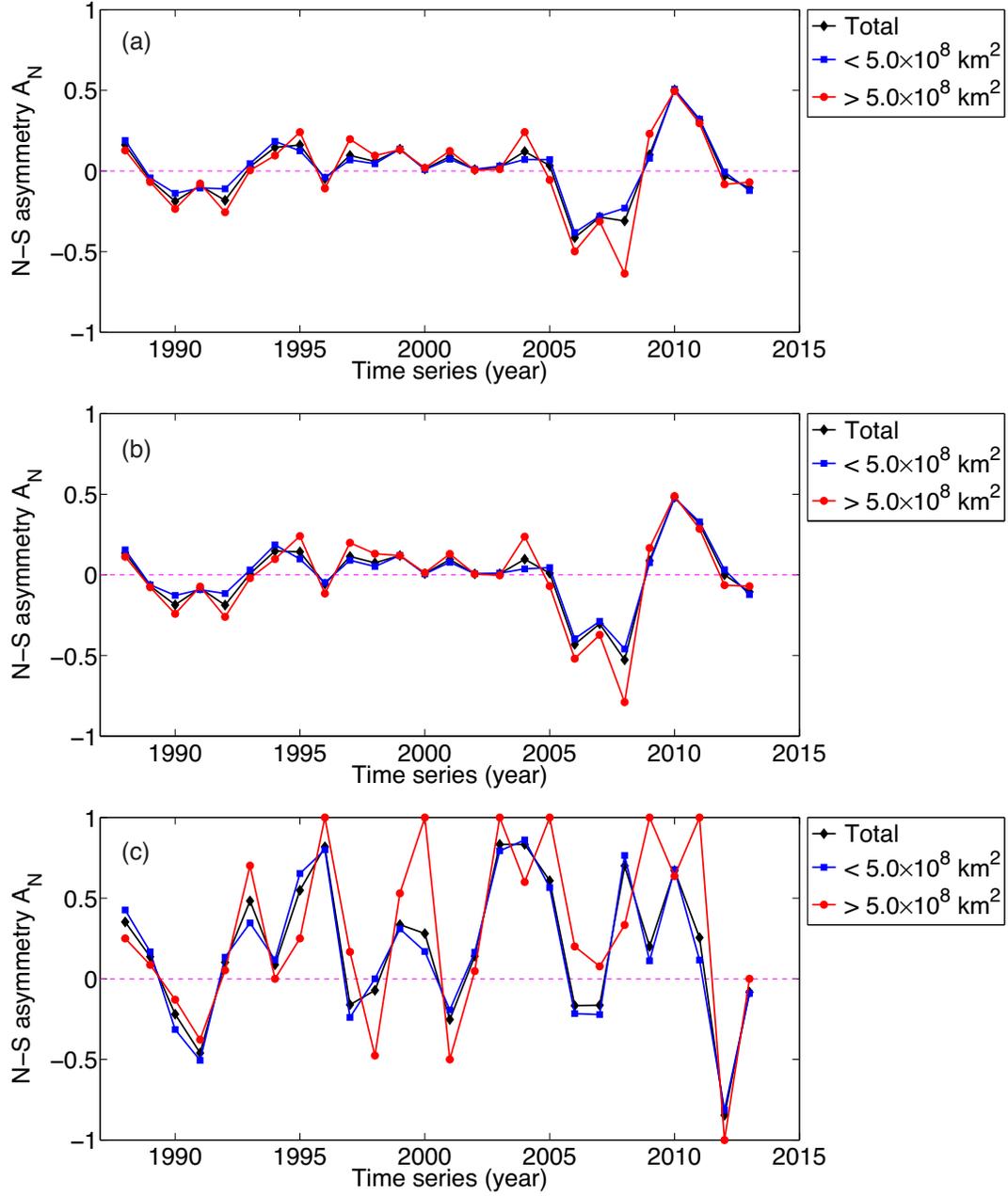}
\caption{The yearly asymmetry of the filament numbers with areas smaller (larger) than $5\times10^{8}$ km$^{2}$ in (a) all latitudes, the latitude bands (b) $0^{\circ}  \sim  50^{\circ}$ and (c) $50^{\circ}  \sim  90^{\circ}$  from 1988 to 2013.
}
\label{fig18}
\end{figure}

\begin{figure}
\centering
\includegraphics[width=1.0\textwidth,clip=]{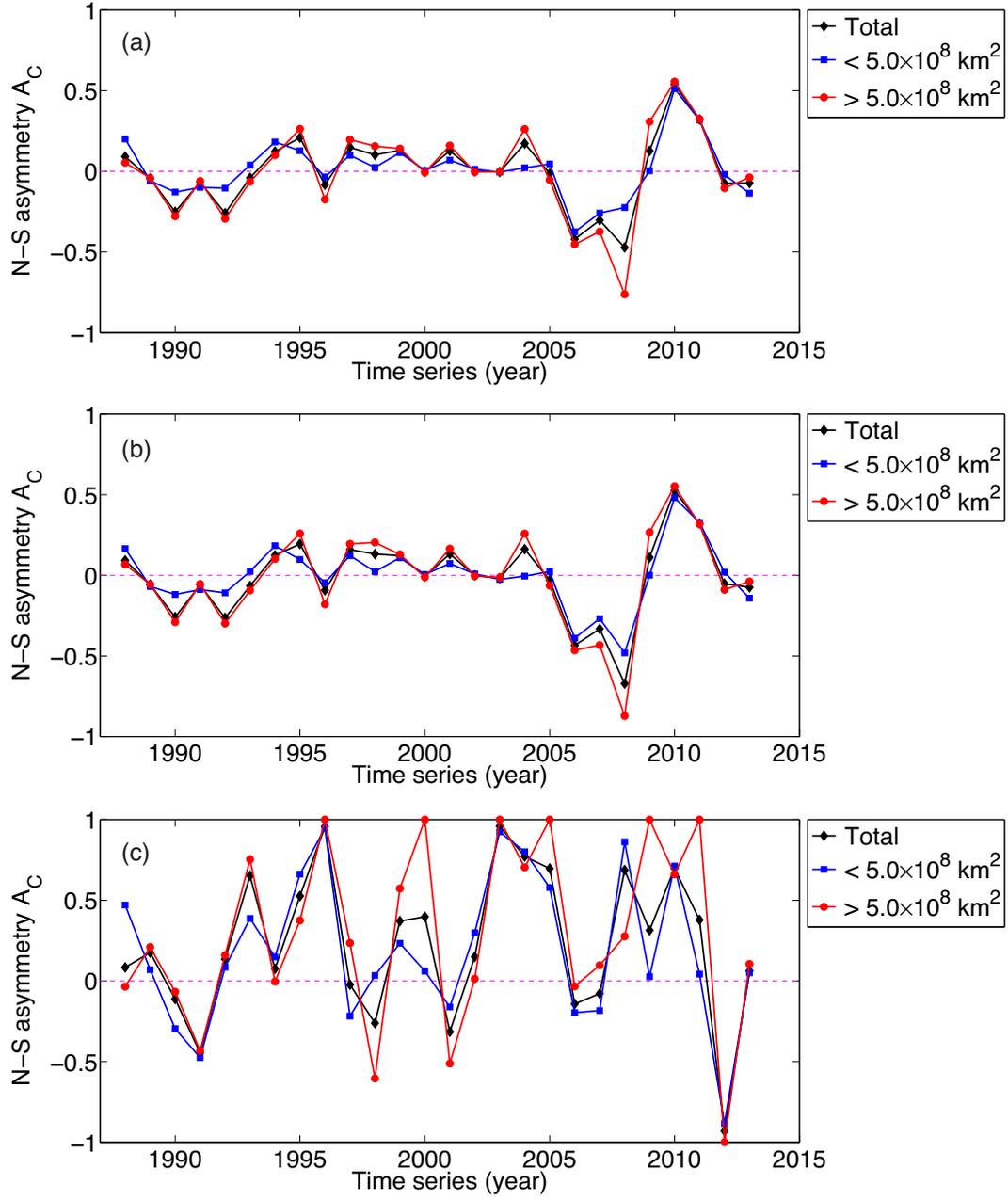}
\caption{Similar to Figure~\ref{fig18}, but for the cumulative filament areas.
}
\label{fig19}
\end{figure}

\begin{figure}
\centering
\includegraphics[width=1.0\textwidth,clip=]{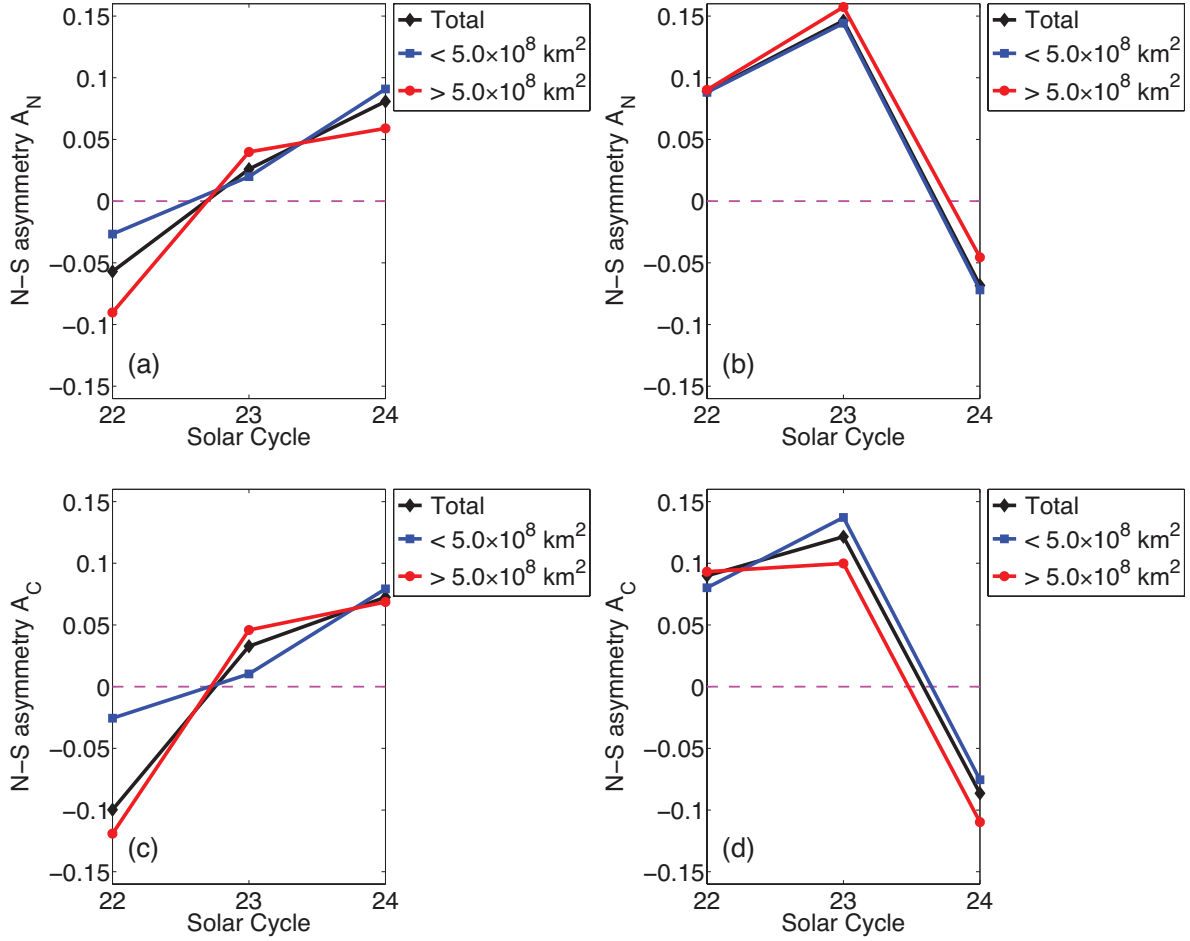}
\caption{The asymmetries of the filament numbers and the cumulative filament areas with areas smaller (larger) than $5\times10^{8}$ km$^{2}$ in the latitude bands $0^{\circ}  \sim  50^{\circ}$ (a), (c) and $50^{\circ}  \sim  90^{\circ}$ (b), (d) in the three solar cycles.
}
\label{fig20}
\end{figure}

\begin{figure}
\centering
\includegraphics[width=1.0\textwidth,clip=]{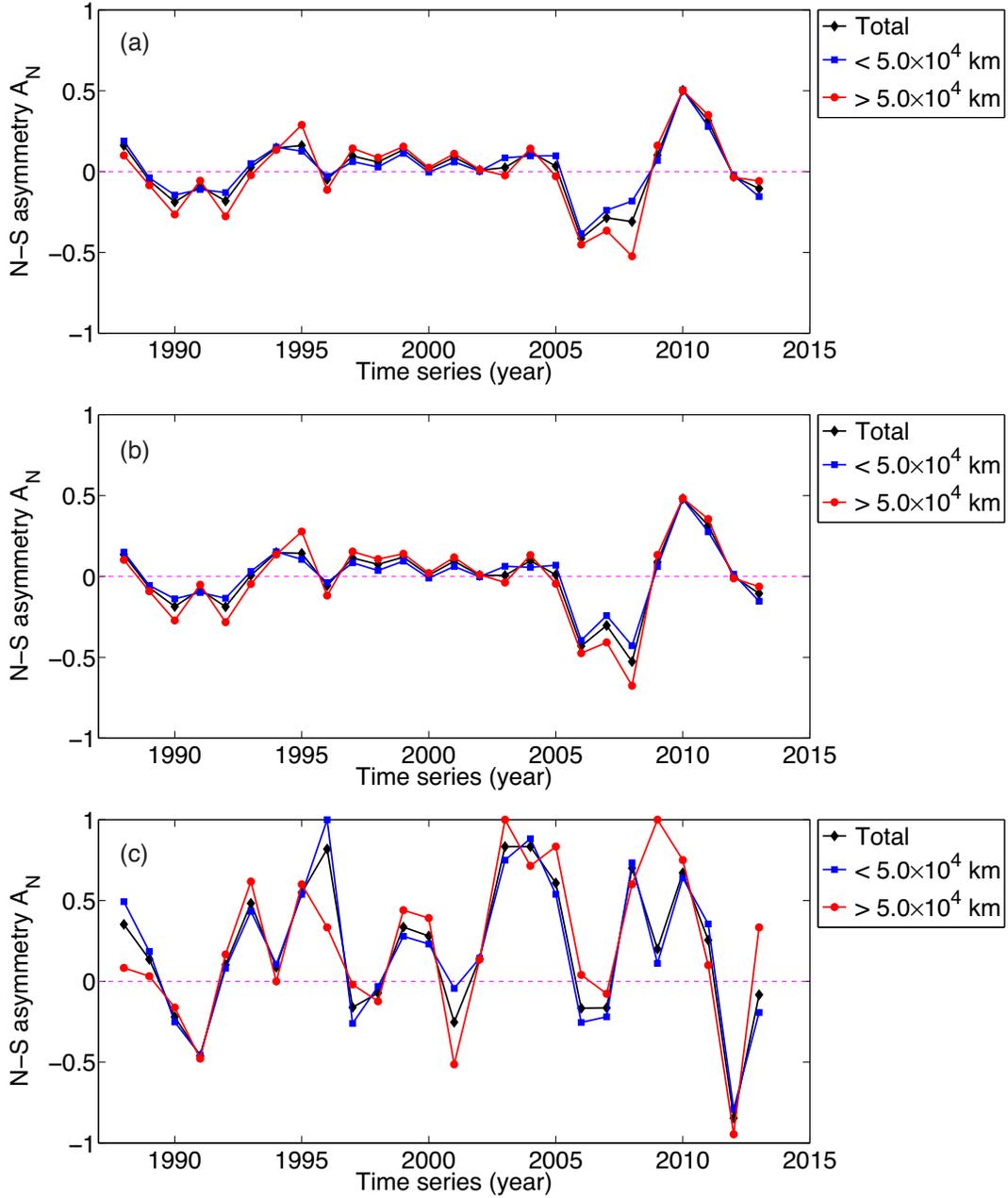}
\caption{Similar to Figure~\ref{fig18}, but for the filament numbers with spine length shorter (longer) than $5\times10^{4}$ km.
}
\label{fig21}
\end{figure}

\begin{figure}
\centering
\includegraphics[width=1.0\textwidth,clip=]{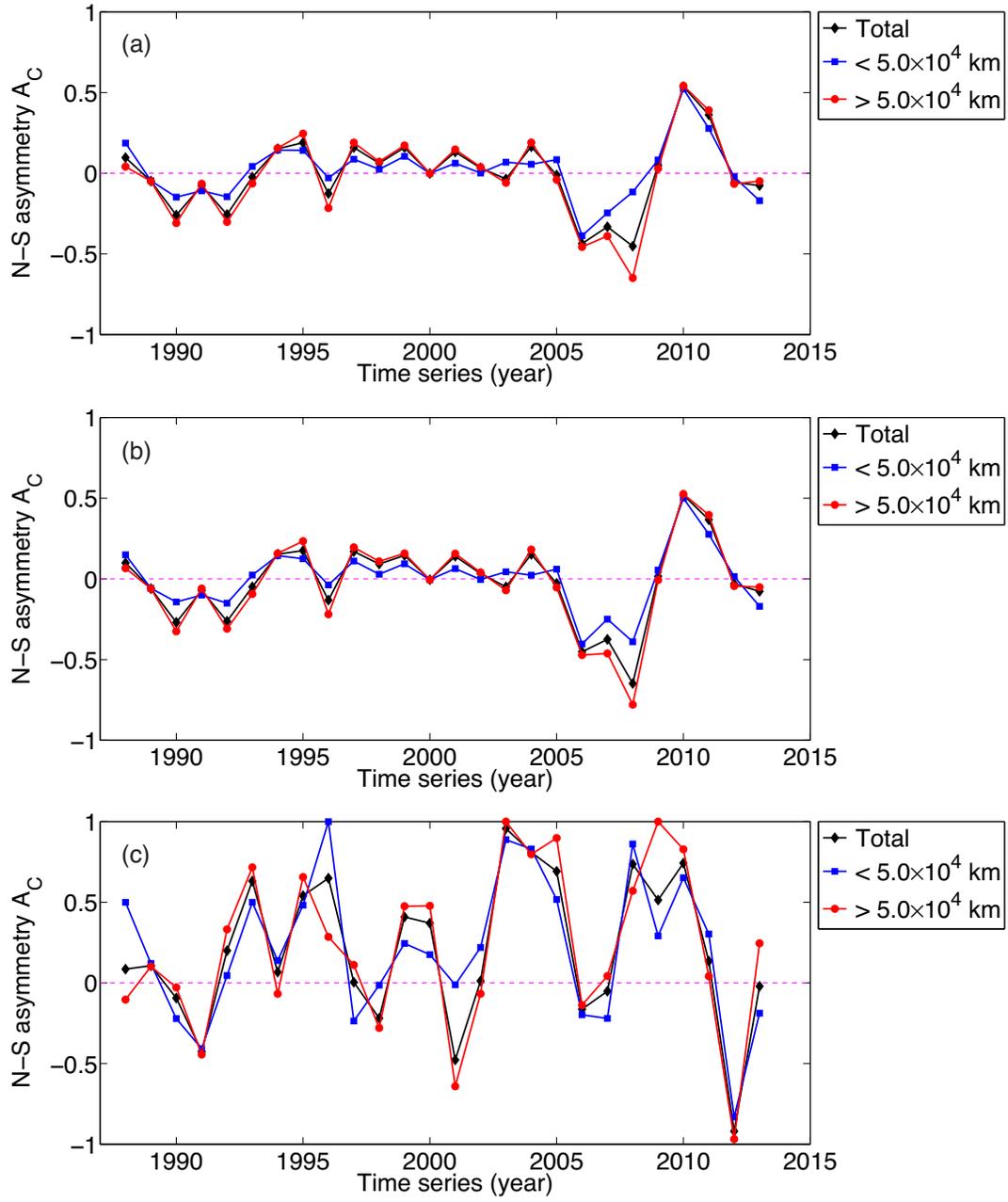}
\caption{Similar to Figures~\ref{fig18} and \ref{fig21}, but for the cumulative spine lengths.
}
\label{fig22}
\end{figure}

\begin{figure}
\centering
\includegraphics[width=1.0\textwidth,clip=]{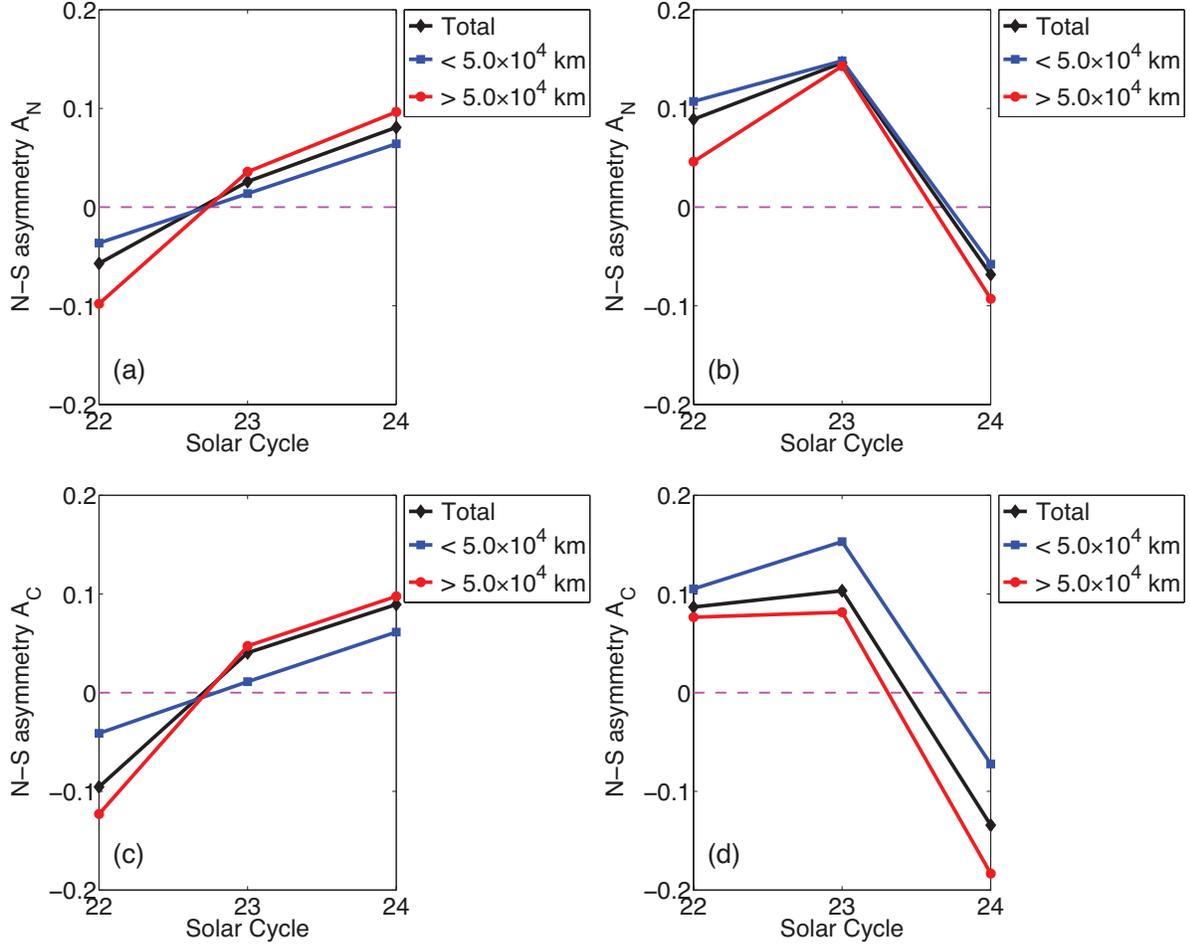}
\caption{Similar to Figure~\ref{fig20}, but for the filament numbers and the cumulative filament spine lengths with spine lengths shorter (longer) than $5\times10^{4}$ km in the latitude bands $0^{\circ}  \sim  50^{\circ}$ (a), (c) and $50^{\circ}  \sim  90^{\circ}$ (b), (d) in the three solar cycles.
}
\label{fig23}
\end{figure}

\begin{figure}
\centering
\includegraphics[width=1.0\textwidth,clip=]{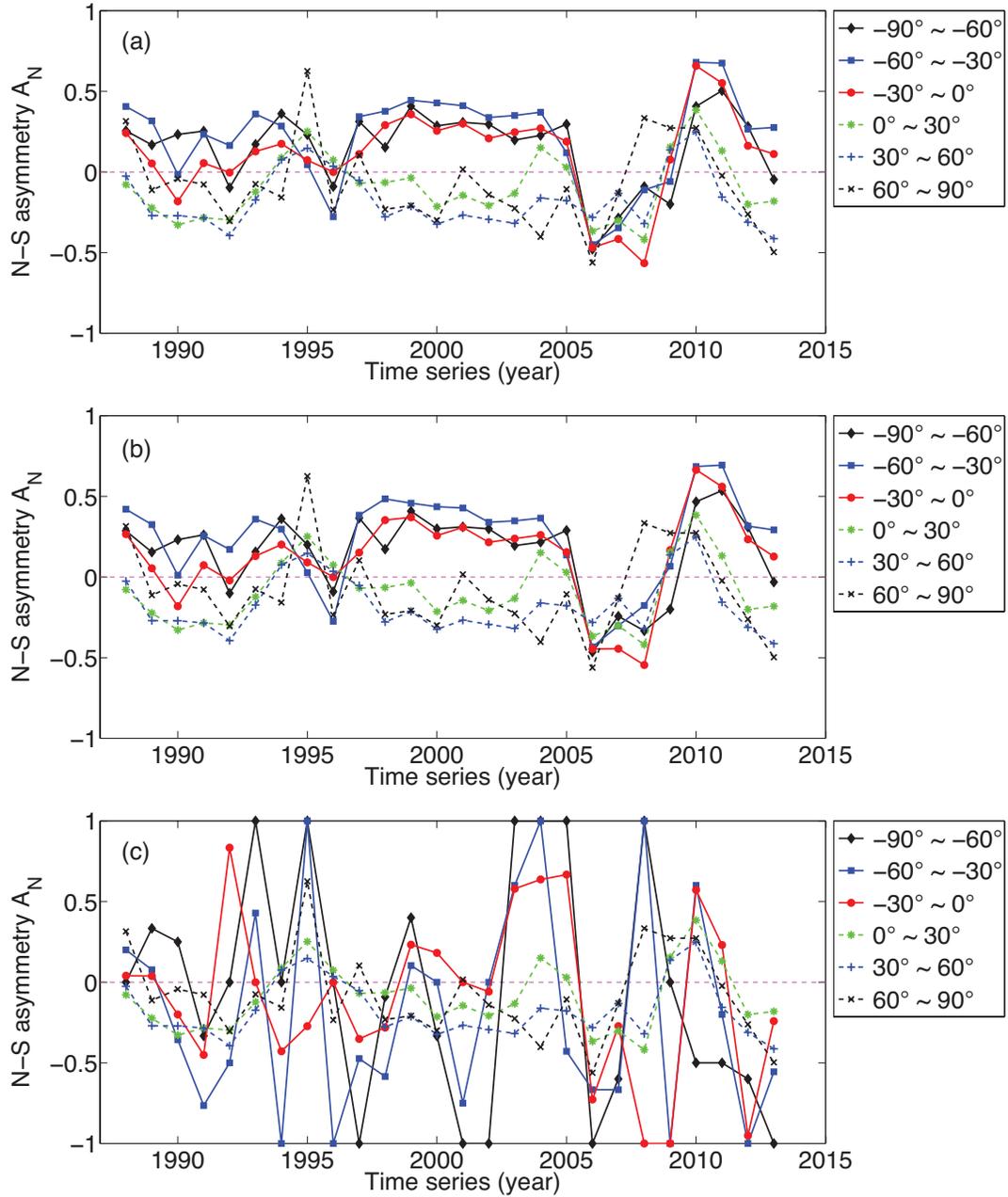}
\caption{Similar to Figure~\ref{fig18}, but for the filament numbers within different tilt angles.
}
\label{fig24}
\end{figure}

\begin{figure}
\centering
\includegraphics[width=1.0\textwidth,clip=]{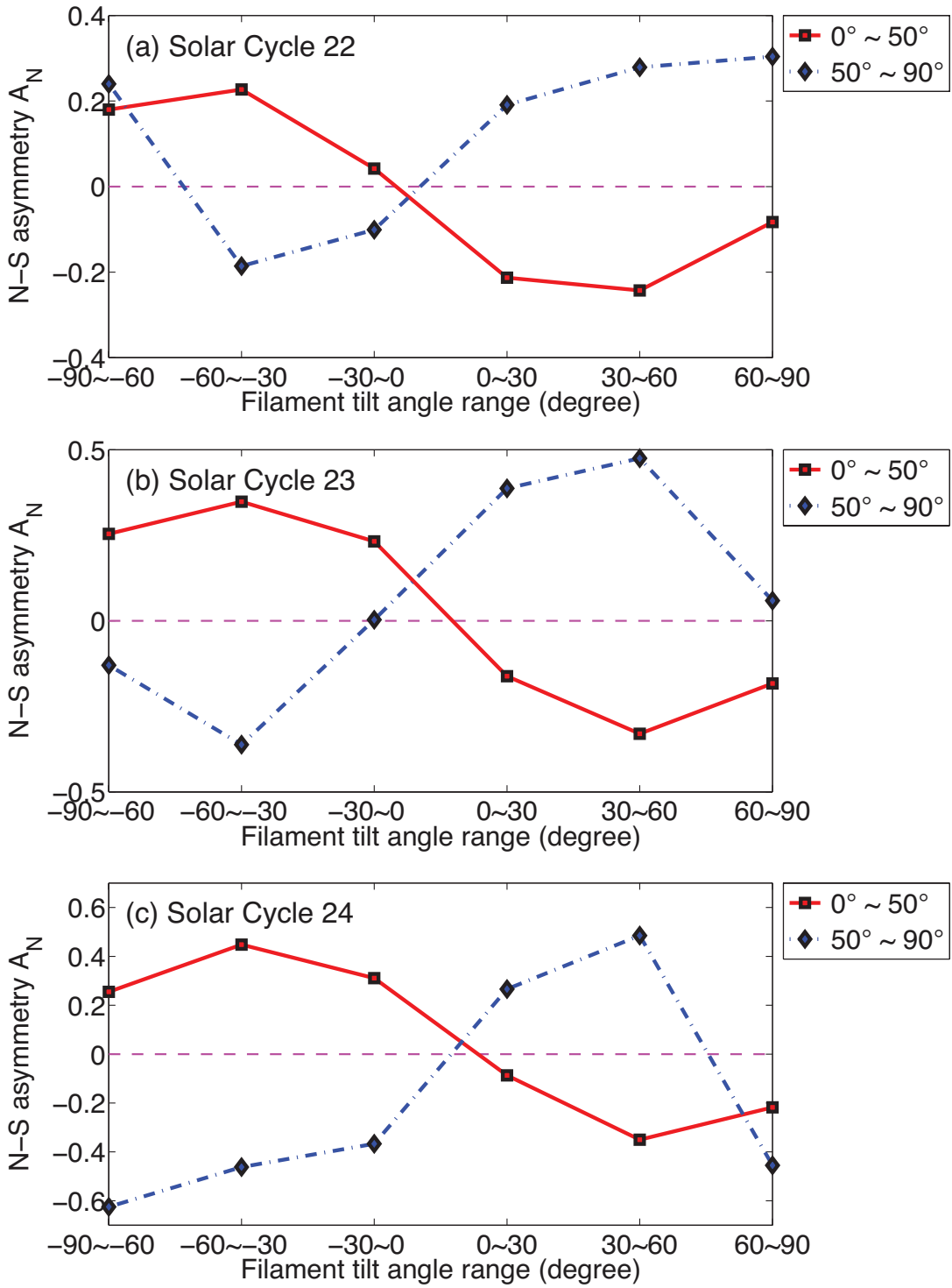}
\caption{ The  asymmetries of the filament numbers with different tilt angles in the latitude bands $0^{\circ}  \sim  50^{\circ}$ and $50^{\circ}  \sim  90^{\circ}$ in the three solar cycles.
}
\label{fig25}
\end{figure}

We employed a normalized N-S asymmetry index to describe the N-S asymmetries of the total filament number and the filament number with different areas, spine lengths, and tilt angles. Besides, we adopted normalized cumulative N-S asymmetry index to describe the N-S asymmetries of the cumulative filament areas and spine lengths. We also calculated the N-S asymmetry indices of the filament features in the latitude bands $0^{\circ}  \sim  50^{\circ}$ and $50^{\circ}  \sim  90^{\circ}$. The yearly N-S asymmetry of the two kind indices are obtained and plotted. Since $90\%$ of all the filaments are located in latitudes lower than $50^{\circ}$, the N-S asymmetry indices in low latitudes and all latitudes have similar trends. The N-S asymmetry of the filament numbers and the cumulative filament areas and spine lengths have similar trends. We found that the N-S asymmetry indices of the filament numbers are all negative in solar cycle 22, which means that the southern hemisphere is dominant, which  is opposite in solar cycle 23, namely, the northern hemisphere is dominant in solar cycle 23. The N-S asymmetry indices in the rising phase of solar cycle 24 indicate that the northern hemisphere is dominant, and the N-S asymmetry tends to strengthen. However, since solar cycle 24 has not finished, here we cannot tell for sure which hemisphere is dominant during solar cycle 24. Our result showing that northern hemisphere is dominant in solar cycle 23 is not consistent with the results given by some other authors, such as \citet{2009Li1} and \citet{2009Joshi}. There may be two reasons for it. One is that their samples are different from ours. For example, \citet{2009Li1} used the data of sunspot groups and sunspot areas, and \citet{2009Joshi} mainly analyzed the active prominences in solar cycle 23. The other one is that each sunspot or prominence in their sample is registered once, whereas in our study the data are not. We detect filaments from one H$\alpha$ image per day, which means that if a filament lasts more than one day it is recorded more than once.  Actually, from the butterfly diagrams in Figures~\ref{fig2}--\ref{fig4}  and the yearly N-S asymmetry indices plotted in Figure~\ref{fig17} we can find that in the period from 2006 to 2008 the southern hemisphere has more filaments than the northern hemisphere. Actually, we found in our study that the N-S asymmetry indices of the total filament numbers in different solar cycles have very small values, which means that the filament numbers in the northern hemisphere and the southern hemisphere do not differ significantly in solar cycles 23. It is noticed that such a small difference exists only among the low latitude filaments. For the high latitude filaments, the difference becomes much larger. \citet{2003Li2} found that the N-S asymmetry of solar activities may be a function of latitude.  Here we cannot give any clear conclusions. The N-S asymmetry indices of the filaments with different areas and the spine lengths show similar results as that of filament numbers. The N-S asymmetry index variations with the filament tilt angles in the latitude band $0^{\circ}  \sim  50^{\circ}$ indicate that the variation curve profiles are similar to the sine functions. The dominant hemisphere is the northern (southern) one when the filament tilt angles within the range $-90^{\circ}  \sim  0^{\circ}$ ($0^{\circ}  \sim  90^{\circ}$).  However, the N-S asymmetry index variations profiles in high latitudes are similar to cosine functions, through some parts are not as clear as that in low latitudes, the dominant orientation of the northern (southern) hemisphere is $0^{\circ}  \sim  90^{\circ}$ ($-90^{\circ}  \sim  0^{\circ}$), which is opposite to that in low latitude bands. These results do not depend on solar cycle.

In summary, we used our improved method to automatically process and analyze 6564 images mainly obtained by BBSO  from April 1988 to December 2013. The variations of the filament features (location, area, spine length, tilt angle, and barb number) with the calendar years and the latitudes were analyzed in detail. The butterfly diagrams of the filaments were obtained and the drift velocities in different latitude bands and the variations with areas and spine lengths were obtained and analyzed. Finally, we investigated the N-S asymmetries of the filament number, their area, spine length and tilt angle. Our main conclusions are as follows:
\\
1) We obtained the butterfly diagrams of the filaments from April 1988 to December 2013 within solar cycles 22, 23, and the rising phase of solar cycle 24.
\\
2) The latitudinal distribution of filament number is bimodal. About $80\%$ of all the filaments are located within  $ 0^{\circ}  \sim  50^{\circ}$ and the peak filament number is within the latitude band $ 10^{\circ}  \sim  30^{\circ}$ in both hemispheres from April 1988 to December 2013 in solar cycles 22, 23, and the rising phase of solar cycle 24.
\\
3) The latitudinal distributions of the filaments with different features (area, spine length, tilt angle and barb number) are also bimodal.
\\
4) Over $90\%$ of all the detected filaments in solar cycle 23 and the rising phase of solar cycle 24 have areas less than $1.0\times10^{9}$ km$^{2}$. The filament spine lengths are mainly less than $1.0\times10^{5}$ km for $85\%$ of all the detected filaments in solar cycle 23 and the rising phase of solar cycle 24. About $80\%$ of all the filaments have tilt angles within $[0^{\circ}, 60^{\circ}]$ in both hemispheres from 1988 to 2013 during the three solar cycles. About $75\%$ filaments have less than 5 barbs in solar cycle 23 and the rising phase of solar cycle 24.
\\
5) For filaments within the latitudes lower than $50^{\circ}$ (namely the low latitude filaments), the northeast and the southeast directions are the dominant orientation in the northern and the southern hemispheres, respectively. Whereas the filaments within the latitude band $ 50^{\circ}  \sim  90^{\circ}$ (namely the high latitude filaments)  the northwest and the southwest directions are the dominant orientation in the northern and the southern hemispheres, respectively.
\\
6) The latitudinal migration of the filaments has three trends in solar cycles 22 and 23: from the beginning of each solar cycle to the solar maximum the drift velocity is high. From the solar maximum to the years before the solar minimum the drift velocity becomes relatively lower. After that the migration becomes divergent. More filaments appear in high latitudes, implying the start of a new cycle.
\\
7) Most filaments in latitudes lower than $50^{\circ}$ migrate toward the equator during the whole solar cycle and most filaments in latitudes higher than $50^{\circ}$ migrate toward the polar region most of the time, especially in the rising phase of each solar cycle.
\\
8) The N-S asymmetry indices of filament numbers, filament numbers with various areas, spine lengths, and the cumulative areas and spine lengths indicate that the southern hemisphere is the dominant one in solar cycle 22 and the northern hemisphere is the dominant one in solar cycle 23, though the difference between two hemispheres is not significant. The N-S asymmetry indices show that the northern hemisphere dominates in the rising phase of solar cycle 24 and has a strengthening trendency. 
\\
9) The N-S asymmetry index variations with the filament tilt angles in the latitude band $0^{\circ}  \sim  50^{\circ}$ shows that the variation profiles are similar to the sine function. The dominant hemisphere is the northern (southern) hemisphere when the filament tilt angles within the range $-90^{\circ}  \sim  0^{\circ}$ ($0^{\circ}  \sim  90^{\circ}$) and these results do not depend on solar cycle.

\begin{acknowledgements}
We thank Big Bear Solar Observatory (BBSO) team and New Jersey Institute of Technology (NJIT) for making the data available. We also thank the referee very much for the constructive suggestions which greatly improved the paper in various ways. This work was supported by NKBRSF under grants 2011CB811402 and 2014CB744203, NSFC grants 11533005, 11203014, and 11025314, as well as the grants CSC201306190046 and CXZZ130041.
\end{acknowledgements}

%\bibliographystyle{apj}
%\bibliography{bibliography_sci}

\clearpage
\begin{table*}  \scriptsize
\centering
\caption{Automatically detected filament numbers from 1988 to 2013.} \label{table1}
\begin{tabular}{lclclclclclclclclclclclclcl} \\
\hline \hline

Time interval   & Solar Cycle  & Data numbers &		&					  	  &Detected    & filament         &	numbers 				    &\\
      		     &                     &                         & Total	&	   &North  	        			  &					       &	  &South    	                     &\\
      		     &                     &                         & 		&Total &$ 0^{\circ}  \sim  50^{\circ}$ &$ 50^{\circ}  \sim  90^{\circ}$  &Total  &$ 0^{\circ}  \sim  50^{\circ}$ &$ 50^{\circ}  \sim  90^{\circ}$   \\
\hline
1988 Apr -- 1996 Aug & 22   &$2,421$  & $26,771$ 	&$12,713$       &$12,034$ 		&$679$  					 &$14,058$   &$13,490$ 	 &$568$   \\
\hline
 1996 Sep -- 2008 Dec   & 23 & $3,124$  & $32,356$    &$16,675$       &$15,938$ 		&$737$  					 &$15,681$   &$15,132$ 	 &$549$   \\
\hline
 2009 Jan -- 2013 Dec & 24 & $1,019$  & $8,305$   	&$4,464$       &$4,314$ 		&$150$  					 &$3,841$   &$3,669$ 	 &$172$   \\
\hline
 1988 Apr -- 2013 Dec & 22-24 &$6,564$  & $67,432$   &$33,852$       &$32,286$ 		&$1,566$  					 &$33,580$   &$32,291$ 	 &$1,289$   \\
\hline
\end{tabular}
\end{table*}

\begin{table*} \scriptsize
\centering
\caption{Automatically detected filament number percentages in different latitude bands and area ranges from 1996 to 2013.} \label{table2}
\begin{tabular}{lclclclclclclclc} \\
\hline \hline

 Solar Cycle  & Latitude band &Filament num & &   Area &$\times10^{8}$(km$^{2}$)&\\
			&			  & 		  &$< 2.5$  &$2.5 \sim 5$  &$5 \sim 10$  &$ >10$\\
\hline
   & $ -90^{\circ}  \sim  0^{\circ}	$&$15,681$    &$33.72\%$ 	&$36.77\%$  	     &$19.46\%$		  &$10.05\%$\\
   & $ -90^{\circ}  \sim  -50^{\circ}	$&$549  $       &$46.45\%$ 	&$38.43\%$  	     &$11.29\%$		  &$3.83\%$\\
   & $ -50^{\circ}  \sim  0^{\circ}	$&$15,132$    &$33.26\%$ 	&$36.71\%$  	     &$19.75\%$		  &$10.28\%$\\
23 & $ 0^{\circ}  \sim  50^{\circ}	$&$15,938$    &$33.50\%$ 	&$35.63\%$  	     &$20.42\%$		  &$10.45\%$\\
   & $ 50^{\circ}  \sim  90^{\circ}	$&$737$         &$49.66\%$ 	&$34.87\%$  	     &$12.21\%$		  &$3.26\%$\\
   & $ 0^{\circ}  \sim  90^{\circ}	$&$16,675$    &$34.21\%$ 	&$35.59\%$  	     &$20.05\%$		  &$10.15\%$\\
   & $ -90^{\circ}  \sim  90^{\circ}	$&$32,356$    &$33.98\%$ 	&$36.16\%$  	     &$19.76\%$		  &$10.10\%$\\
\hline
  & $ -90^{\circ}  \sim  0^{\circ}	$&$3,841$    &$28.56\%$ 	&$39.89\%$  	     &$20.83\%$		  &$10.72\%$\\
   & $ -90^{\circ}  \sim  -50^{\circ}	$&$172   $    &$33.14\%$ 	&$53.49\%$  	     &$9.88\%$		  &$3.49\%$\\
   & $ -50^{\circ}  \sim  0^{\circ}	$&$3,669$    &$28.35\%$ 	&$39.25\%$  	     &$21.34\%$		  &$11.06\%$\\
24 & $ 0^{\circ}  \sim  50^{\circ}	$&$4,314$    &$30.00\%$ 	&$38.99\%$  	     &$19.73\%$		  &$11.28\%$\\
   & $ 50^{\circ}  \sim  90^{\circ}	$&$ 150$      &$36.67\%$ 	&$49.33\%$  	     &$12.67\%$		  &$1.33\%$\\
   & $ 0^{\circ}  \sim  90^{\circ}	$&$4,464$    &$30.22\%$ 	&$39.34\%$  	     &$19.49\%$		  &$10.95\%$\\
   & $ -90^{\circ}  \sim  90^{\circ}	$&$8,305$    &$29.45\%$ 	&$39.59\%$  	     &$20.11\%$		  &$10.85\%$\\
\hline
\end{tabular}
\end{table*}

\begin{table*}  \scriptsize
\centering
\caption{Automatically detected filament number percentages in different latitude bands and spine length ranges from 1996 to 2013.} \label{table3}
\begin{tabular}{lclclclclclclclc} \\
\hline \hline

 Solar Cycle  & Latitude band &Filament num &  &Spine length &  $\times10^{4}$(km)&&\\
			&			  & 		  &$< 5$  &$5 \sim 10$  &$>10 $  \\
\hline
   & $ -90^{\circ}  \sim  0^{\circ}	$&$15,681$    &$45.83\%$ 	     &$30.94\%$		  &$23.23\%$\\
   & $ -90^{\circ}  \sim  -50^{\circ}	$&$549  $       &$62.84\%$	     &$22.40\%$		  &$14.76\%$\\
   & $ -50^{\circ}  \sim  0^{\circ}	$&$15,132$    &$45.21\%$ 	     &$31.25\%$		  &$23.54\%$\\
23 & $ 0^{\circ}  \sim  50^{\circ}	$&$15,938$    &$44.11\%$      &$31.88\%$		  &$24.01\%$\\
   & $ 50^{\circ}  \sim  90^{\circ}	$&$737  $       &$63.09\%$      &$23.34\%$		  &$13.57\%$\\
   & $ 0^{\circ}  \sim  90^{\circ}	$&$16,675$    &$44.95\%$      &$31.50\%$		  &$23.55\%$\\
   & $ -90^{\circ}  \sim  90^{\circ}	$&$32,356$    &$45.38\%$      &$31.23\%$		  &$23.39\%$\\
\hline
  & $ -90^{\circ}  \sim  0^{\circ}	$&$3,841$    &$50.17\%$ 	     &$30.88\%$		  &$18.95\%$\\
   & $ -90^{\circ}  \sim  -50^{\circ}	$&$172   $    &$69.19\%$  	     &$20.93\%$		  &$9.88\%$\\
   & $ -50^{\circ}  \sim  0^{\circ}	$&$3,669$    &$49.28\%$  	     &$31.34\%$		  &$19.38\%$\\
24 & $ 0^{\circ}  \sim  50^{\circ}	$&$4,314$    &$47.66\%$  	     &$31.66\%$		  &$20.68\%$\\
   & $ 50^{\circ}  \sim  90^{\circ}	$&$  150$    &$70.67\%$   	     &$24.00\%$		  &$5.33\%$\\
   & $ 0^{\circ}  \sim  90^{\circ}	$&$4,464$    &$48.43\%$   	     &$31.41\%$		  &$20.16\%$\\
   & $ -90^{\circ}  \sim  90^{\circ}	$&$8,305$    &$49.24\%$  	     &$31.16\%$		  &$19.60\%$\\
\hline
\end{tabular}
\end{table*}

\begin{table*} \scriptsize
\centering
\caption{Automatically detected filament number percentages in different latitude bands and tilt angle directions from 1988 to 2013.} \label{table4}
\begin{tabular}{lclclclclclclc} \\
\hline \hline
 Solar Cycle  & Direction & &  &&Latitude band &&\\
			&			 &$ -90^{\circ}  \sim  0^{\circ}$  &$-90^{\circ}  \sim  -50^{\circ}$  &$ -50^{\circ}  \sim  0^{\circ}$  &$0^{\circ}  \sim  50^{\circ}$  &$50^{\circ}  \sim  90^{\circ}$  &$0^{\circ}  \sim  90^{\circ}$  &$-90^{\circ}  \sim  90^{\circ} $ \\
\hline
  22 & NW or SE$^{\mathrm{a}}$ &$61.74\%$ 	&$49.82\%$  	     &$62.27\%$		  &$46.21\%$		 &$67.50\%$		       &$47.40\%$		    &$54.88\%$\\
      & NE or SW$^{\mathrm{b}}$ &$38.26\%$ 	&$50.18\%$  	     &$37.73\%$		  &$53.79\%$		 &$32.50\%$		       &$52.60\%$		    &$45.12\%$\\
\hline
  23 & NW or SE &$62.43\%$ 	&$37.16\%$  	     &$63.38\%$		  &$37.40\%$		&$64.99\%$		       &$38.66\%$		    &$50.19\%$\\
      & NE or SW&$37.57\%$ 	&$62.84\%$  	     &$36.62\%$		  &$62.60\%$		&$35.01\%$		       &$61.34\%$		    &$49.81\%$\\
\hline
  24 & NW or SE &$62.66\%$ 	&$34.88\%$  	     &$63.98\%$		  &$35.99\%$		&$66.00\%$		       &$37.03\%$		    &$48.98\%$\\
      & NE or SW &$37.34\%$ 	&$65.12\%$  	     &$36.02\%$		  &$64.01\%$		&$34.00\%$		       &$62.97\%$		    &$51.02\%$\\
\hline
\end{tabular}

\begin{list}{}{}
\item[$^{\mathrm{a}}$] NW: the filament is northwest direction and locates in the northern hemisphere; SE: the filament is southeast direction and locates in the southern hemisphere.
\item[$^{\mathrm{b}}$] NE: the filament is northeast direction and locates in the northern hemisphere; SW: the filament is southwest direction and locates in the southern hemisphere.
\end{list}
\end{table*}

\begin{table*} \scriptsize
\centering
\caption{Automatically detected filament number percentages in different latitude bands and tilt angle intervals from 1988 to 2013.} \label{table5}
\begin{tabular}{lclclclclclclclc} \\
\hline \hline
 Solar Cycle  & Latitude band &Filament num & &  &Tilt angle &  &&\\
			&			  & 		  &$-90^{\circ}  \sim  -60^{\circ}$  &$-60^{\circ}  \sim  -30^{\circ}$  &$-30^{\circ}  \sim  0^{\circ}$  &$0^{\circ}  \sim  30^{\circ}$  &$30^{\circ}  \sim  60^{\circ}$  &$60^{\circ}  \sim  90^{\circ}$ \\
\hline
   & $ -90^{\circ}  \sim  0^{\circ}	$&$14,058$    &$3.50\%$ 	&$12.34\%$  	     &$24.23\%$		  &$29.33\%$		 &$26.76\%$		       &$3.84\%$		    \\
   & $ -90^{\circ}  \sim  -50^{\circ}	$&$568  $       &$3.35\%$ 	&$14.61\%$  	     &$34.51\%$		  &$29.05\%$		 &$15.67\%$		       &$2.81\%$		    \\
   & $ -50^{\circ}  \sim  0^{\circ}	$&$13,490$    &$3.51\%$ 	&$12.25\%$  	     &$23.80\%$		  &$29.34\%$		 &$27.23\%$		       &$3.87\%$		    \\
22 & $ 0^{\circ}  \sim  50^{\circ}	$&$12,034$    &$5.65\%$ 	&$21.77\%$  	     &$28.99\%$		  &$21.32\%$		 &$18.57\%$		       &$3.70\%$		    \\
   & $ 50^{\circ}  \sim  90^{\circ}	$&$679$         &$4.57\%$ 	&$8.39\%$  	     &$23.56\%$		  &$35.79\%$		 &$23.27\%$		       &$4.42\%$		    \\
  & $ 0^{\circ}  \sim  90^{\circ}	$&$12,713$    &$5.59\%$ 	&$21.06\%$  	     &$28.70\%$		  &$22.10\%$		 &$18.82\%$		       &$3.73\%$		    \\
   & $ -90^{\circ}  \sim  90^{\circ}	$&$26,771$    &$4.49\%$ 	&$16.48\%$  	     &$26.35\%$		  &$25.89\%$		 &$22.99\%$		       &$3.80\%$		    \\
\hline
   & $ -90^{\circ}  \sim  0^{\circ}	$&$15,681$    &$5.24\%$ 	&$13.65\%$  	     &$19.00\%$		  &$28.27\%$		 &$26.08\%$		       &$7.76\%$		    \\
   & $ -90^{\circ}  \sim  -50^{\circ}	$&$549  $       &$4.74\%$ 	&$22.95\%$  	     &$34.61\%$		  &$24.77\%$		 &$8.56\%$		       &$4.37\%$		    \\
   & $ -50^{\circ}  \sim  0^{\circ}	$&$15,132$    &$5.25\%$ 	&$13.31\%$  	     &$18.44\%$		  &$28.40\%$		 &$26.72\%$		       &$7.88\%$		    \\
23 & $ 0^{\circ}  \sim  50^{\circ}	$&$15,938$    &$8.38\%$ 	&$26.13\%$  	     &$28.10\%$		  &$19.45\%$		 &$12.77\%$		       &$5.17\%$		    \\
  & $ 50^{\circ}  \sim  90^{\circ}	$&$737$         &$2.71\%$ 	&$8.01\%$  	     &$25.92\%$		  &$41.79\%$		 &$17.91\%$		       &$3.66\%$		    \\
   & $ 0^{\circ}  \sim  90^{\circ}	$&$16,675$    &$8.13\%$ 	&$25.33\%$  	     &$28.00\%$		  &$20.44\%$		 &$13.00\%$		       &$5.10\%$		    \\
   & $ -90^{\circ}  \sim  90^{\circ}	$&$32,356$    &$6.73\%$ 	&$19.67\%$  	     &$23.64\%$		  &$24.23\%$		 &$19.34\%$		       &$6.39\%$		    \\
\hline
   & $ -90^{\circ}  \sim  0^{\circ}	$&$3,841$    &$5.62\%$ 	&$12.47\%$  	     &$19.14\%$		  &$26.27\%$		 &$27.28\%$		       &$9.22\%$		    \\
   & $ -90^{\circ}  \sim  -50^{\circ}	$&$172  $     &$7.56\%$ 	&$22.09\%$  	     &$38.95\%$		  &$16.86\%$		 &$9.88\%$		       &$4.66\%$		    \\
   & $ -50^{\circ}  \sim  0^{\circ}	$&$3,669$    &$5.53\%$ 	&$12.02\%$  	     &$18.21\%$		  &$26.71\%$		 &$28.10\%$		       &$9.43\%$		    \\
24 & $ 0^{\circ}  \sim  50^{\circ}	$&$4,314$    &$7.93\%$ 	&$26.84\%$  	     &$29.46\%$		  &$19.10\%$		 &$11.52\%$		       &$5.15\%$		    \\
   & $ 50^{\circ}  \sim  90^{\circ}	$&$150  $     &$2.00\%$ 	&$9.33\%$  	     &$20.67\%$		  &$33.33\%$		 &$32.67\%$		       &$2.00\%$		    \\
   & $ 0^{\circ}  \sim  90^{\circ}	$&$4,464$    &$7.73\%$ 	&$26.25\%$  	     &$29.17\%$		  &$19.58\%$		 &$12.23\%$		       &$5.04\%$		    \\
   & $ -90^{\circ}  \sim  90^{\circ}	$&$8,305$    &$6.75\%$ 	&$19.88\%$  	     &$24.53\%$		  &$22.67\%$		 &$19.19\%$		       &$6.98\%$		    \\
\hline
\end{tabular}
\end{table*}

\begin{table*} \scriptsize
\centering
\caption{Automatically detected filament number percentages in different latitude bands and barb number intervals from 1996 to 2013.} \label{table6}
\begin{tabular}{lclclclclclclclc} \\
\hline \hline
 Solar Cycle  & Latitude band &Filament num &  &Barb num &  &&\\
			&			  & 		  &$0  \sim  5$  &$6  \sim  10$    &$> 10 $ \\
\hline
   & $ -90^{\circ}  \sim  0^{\circ}	$&$15,681$    &$76.65\%$ 	&$17.72\%$  	     &$5.63\%$\\
   & $ -90^{\circ}  \sim  -50^{\circ}	$&$549$         &$80.69\%$ 	&$14.75\%$  	     &$4.56\%$\\
   & $ -50^{\circ}  \sim  0^{\circ}	$&$15,132$    &$76.50\%$ 	&$17.82\%$  	     &$5.68\%$\\
23 & $ 0^{\circ}  \sim  50^{\circ}	$&$15,938$    &$74.79\%$ 	&$18.76\%$  	     &$6.45\%$\\
   & $ 50^{\circ}  \sim  90^{\circ}	$&$737$         &$79.78\%$ 	&$16.69\%$  	     &$3.53\%$\\
   & $ 0^{\circ}  \sim  90^{\circ}	$&$16,675$    &$75.01\%$ 	&$18.67\%$  	     &$6.32\%$\\
   & $ -90^{\circ}  \sim  90^{\circ}	$&$32,356$    &$75.80\%$ 	&$18.21\%$  	     &$5.99\%$\\
\hline
   & $ -90^{\circ}  \sim  0^{\circ}	$&$3,841$    &$74.88\%$ 	&$20.07\%$  	     &$5.05\%$\\
   & $ -90^{\circ}  \sim  -50^{\circ}	$&$172  $    &$85.47\%$ 	&$13.37\%$  	     &$1.16\%$ \\
   & $ -50^{\circ}  \sim  0^{\circ}	$&$3,669$    &$74.38\%$ 	&$20.39\%$  	     &$5.23\%$\\
24 & $ 0^{\circ}  \sim  50^{\circ}	$&$4,314$    &$73.81\%$ 	&$20.70\%$  	     &$5.49\%$\\
   & $ 50^{\circ}  \sim  90^{\circ}	$&$150  $    &$74.67\%$ 	&$23.33\%$  	     &$2.00\%$\\
   & $ 0^{\circ}  \sim  90^{\circ}	$&$4,464$    &$73.84\%$ 	&$20.79\%$  	     &$5.37\%$ \\
   & $ -90^{\circ}  \sim  90^{\circ}	$&$8,305$    &$74.32\%$ 	&$20.46\%$  	     &$5.22\%$\\
\hline
\end{tabular}
\end{table*}

\begin{table*} \scriptsize
\centering
\caption{Asymmetry of the automatically detected filament numbers in the northern and southern hemispheres from 1988 to 2013.} \label{table7}
\begin{tabular}{lclclclclclclclc} \\
\hline \hline
Time interval   		& Solar Cycle  &Latitude band 				&$N_{N}$		&$N_{S}$			 &$A_{N}$    &Dominant hemisphere\\
\hline
1988 Apr -- 1996 May & 22   	     &$0^{\circ}  \sim  90^{\circ}$  		& $12,713$ 	&$14,058$       		 &$-0.050$ 		 &South 		 \\
				&	  	     &$0^{\circ}  \sim  50^{\circ}$  		& $12,034$ 	&$13,490$       		 &$-0.057$ 		 &South		 \\
				&	  	     &$50^{\circ}  \sim  90^{\circ}$  	& $679$ 		&$568$       		 &$0.089$ 			 &North		 \\
\hline
 1996 Jun -- 2008 Jan & 23 	     &$0^{\circ}  \sim  90^{\circ}$	  	& $16,675$    	&$15,681$       		 &$0.031$ 		 	&North  		 \\
				&	  	     &$0^{\circ}  \sim  50^{\circ}$  		& $15,938$ 	&$15,132$       		 &$0.026$ 		 	&North		 \\
				&	  	     &$50^{\circ}  \sim  90^{\circ}$  	& $737$ 		&$549$       		 &$0.146$ 			&North		 \\
\hline
 2008 Feb -- 2013 Dec & 24 	     &$0^{\circ}  \sim  90^{\circ}$  		& $4,464$   	&$3,841$      		 &$0.075$ 		 	&North  		 \\
				&	  	     &$0^{\circ}  \sim  50^{\circ}$  		& $4,314$ 		&$3,669$       		 &$0.081$ 		 	&North		 \\
				&	  	     &$50^{\circ}  \sim  90^{\circ}$  	& $150$ 		&$172$       		 &$-0.068$ 		&South		 \\
\hline
\end{tabular}
\end{table*}

\clearpage

\begin{table*} \scriptsize
\centering
\caption{Asymmetry of the filament numbers with different area ranges in the latitude bands $0^{\circ}  \sim  50^{\circ}$  and $50^{\circ}  \sim  90^{\circ}$ in the northern and southern hemispheres from 1988 to 2013.} \label{table8}
\begin{tabular}{lclclclclc} \\
\hline \hline
&&Low latitude bands     &($0^{\circ}  \sim  50^{\circ}$)\\			
Time interval   		& Solar Cycle  &Area ($\times10^{8} $ km$^{2}$)		&$N_{N}$		&$S_{N}$			 &$A_{N}$    &Dominant hemisphere\\
\hline
1988 Apr -- 1996 May & 22   	     &$< 5$ 		& $6,489$ 		&$6,845$       		 &$-0.027$ 		&South 		 \\
				&	  	     &$> 5$   	& $5,545$ 		&$6,645$       		 &$-0.090$ 		&South		 \\
\hline
 1996 Jun -- 2008 Jan & 23 	     &$<5$	  	& $11,017$    	&$10,588$       		 &$0.020$ 			&North  		 \\
				&	  	     &$>5$   	& $4,921$ 		&$4,544$       		 &$0.040$ 			&North		 \\
\hline
 2008 Feb -- 2013 Dec & 24 	     &$<5$  		& $2,976$   	&$2,480$      		 &$0.091$ 			&North  		 \\
				&	  	     &$>5$   	& $1,338$ 		&$1,189$       		 &$0.059$ 			&North		 \\
\hline \hline
&&High latitude bands     &($50^{\circ}  \sim  90^{\circ}$)\\	
Time interval   		& Solar Cycle  &Area ($\times10^{8} $ km$^{2}$)		&$N_{N}$		&$S_{N}$			 &$A_{N}$    &Dominant hemisphere\\
\hline
1988 Apr -- 1996 May & 22   	     &$<5$ 		& $401$ 		&$336$       		 &$0.088$ 			&North 		 \\
				&	  	     &$>5$   	& $278$ 		&$232$       		 &$0.090$ 			&North		 \\
\hline
 1996 Jun -- 2008 Jan & 23 	     &$<5$	  	& $623$    		&$466$       		 &$0.144$ 			&North  		 \\
				&	  	     &$>5$   	& $114$ 		&$83$       		 	 &$0.157$ 			&North  		 \\
\hline
 2008 Feb -- 2013 Dec & 24 	     &$<5$  		& $129$   		&$149$      		 &$-0.072$ 		&South  		 \\
				&	  	     &$>5$   	& $21$ 		&$23$       		 	 &$-0.046$ 		&South		 \\
\hline
\end{tabular}
\end{table*}

\begin{table*} \scriptsize
\centering
\caption{Asymmetry of the cumulative filament areas with different area ranges in the latitude bands $0^{\circ}  \sim  50^{\circ}$  and $50^{\circ}  \sim  90^{\circ}$ in the northern and southern hemispheres from 1988 to 2013.} \label{table9}
\begin{tabular}{lclclclclc} \\
\hline \hline
&&Low latitude bands     &($0^{\circ}  \sim  50^{\circ}$)\\			
Time interval   		& Solar Cycle  &Area ($\times10^{8} $ km$^{2}$)		&$N_{C}$ (km$^{2}$)			&$S_{C}$ (km$^{2}$)			 &$A_{C}$    &Dominant hemisphere\\
\hline
1988 Apr -- 1996 May & 22   	     &$< 5$ 		& $2.073\times10^{12}$ 		&$2.182\times10^{12}$       		 &$-0.026$ 		&South 		 \\
				&	  	     &$> 5$   	& $7.167\times10^{12}$ 		&$9.105\times10^{12}$       		 &$-0.119$ 		&South		 \\
\hline
 1996 Jun -- 2008 Jan & 23 	     &$<5$	  	& $2.853\times10^{12}$    		&$2.795\times10^{12}$       		 &$0.010$ 			&North  		 \\
				&	  	     &$>5$   	& $5.051\times10^{12}$ 		&$4.608\times10^{12}$       		 &$0.046$ 			&North		 \\
\hline
 2008 Feb -- 2013 Dec & 24 	     &$<5$  		& $7.943\times10^{11}$   		&$6.776\times10^{11}$      		 &$0.079$ 			&North  		 \\
				&	  	     &$>5$   	& $1.404\times10^{12}$ 		&$1.224\times10^{12}$       		 &$0.069$ 			&North		 \\
\hline \hline
&&High latitude bands     &($50^{\circ}  \sim  90^{\circ}$)\\	
Time interval   		& Solar Cycle  &Area ($\times10^{8} $ km$^{2}$)		&$N_{C}$ (km$^{2}$)		&$S_{C} $ (km$^{2}$)				 &$A_{C}$    &Dominant hemisphere\\
\hline
1988 Apr -- 1996 May & 22   	     &$<5$ 		& $1.183\times10^{11}$ 		&$1.008\times10^{11}$       		 &$0.080$ 			&North 		 \\
				&	  	     &$>5$   	& $3.528\times10^{11}$ 		&$2.927\times10^{11}$       		 &$0.093$ 			&North		 \\
\hline
 1996 Jun -- 2008 Jan & 23 	     &$<5$	  	& $1.295\times10^{11}$    		&$9.822\times10^{10}$       		 &$0.137$ 			&North  		 \\
				&	  	     &$>5$   	& $9.042\times10^{10}$ 		&$7.400\times10^{10}$       		 &$0.100$ 			&North  		 \\
\hline
 2008 Feb -- 2013 Dec & 24 	     &$<5$  		& $3.239\times10^{10}$   		&$3.768\times10^{10}$      		 &$-0.075$ 		&South  		 \\
				&	  	     &$>5$   	& $1.471\times10^{10}$ 		&$1.836\times10^{10}$       		 &$-0.110$ 		&South		 \\
\hline
\end{tabular}
\end{table*}

\begin{table*} \scriptsize
\centering
\caption{Asymmetry of the filament numbers with different spine length ranges in the latitude bands $0^{\circ}  \sim  50^{\circ}$  and $50^{\circ}  \sim  90^{\circ}$ in the northern and southern hemispheres from 1988 to 2013.} \label{table10}
\begin{tabular}{lclclclclc} \\
\hline \hline
&&Low latitude bands     &($0^{\circ}  \sim  50^{\circ}$)\\	
Time interval   		& Solar Cycle  &Spine length ($\times10^{4} $ km)		&$N_{N}$		&$S_{N}$			 &$A_{N}$    &Dominant hemisphere\\
\hline
1988 Apr -- 1996 May & 22   	     &$<5$ 				& $8,167$ 		&$8,784$       		 &$-0.036$ 		&South 		 \\
				&	  	     &$>5$   			& $3,867$ 		&$4,706$       		 &$-0.098$ 		&South		 \\
\hline
 1996 Jun -- 2008 Jan & 23 	     &$<5$	  			& $7,031$    	&$6,841$       		 &$0.014$ 			&North  		 \\
				&	  	     &$>5$   			& $8,907$ 		&$8,291$       		 &$0.036$ 			&North 		 \\
\hline
 2008 Feb -- 2013 Dec & 24             &$<5$                      	& $2,056$           &$1,808$                   &$0.064$                	&North             \\
	                          &                    &$>5$           			& $2,258$           &$1,861$                   &$0.096$                	&North            \\
\hline\hline
&&High latitude bands     &($50^{\circ}  \sim  90^{\circ}$)\\
Time interval   		& Solar Cycle  &Spine length ($\times10^{4} $ km)		&$N_{N}$		&$S_{N}$			 &$A_{N}$    &Dominant hemisphere\\
\hline
1988 Apr -- 1996 May & 22   	     &$<5$ 				& $486$ 		&$392$       		 &$0.107$ 			&North 		 \\
				&	  	     &$>5$   			& $193$ 		&$176$       		 &$0.046$ 			&North		 \\
\hline
 1996 Jun -- 2008 Jan & 23 	     &$<5$	  			& $465$    		&$345$       		 &$0.148$ 			&North  		 \\
				&	  	     &$>5$   			& $272$ 		&$204$       		 &$0.143$ 			&North 		 \\
\hline
 2008 Feb -- 2013 Dec & 24              &$<5$                      	& $106$           &$119$                   	&$-0.058$                	&South             \\
	                          &                     &$>5$           		& $44$             &$53$                   	&$-0.093$                	&South            \\	
\hline
\end{tabular}
\end{table*}

\begin{table*} \scriptsize
\centering
\caption{Asymmetry of the cumulative filament spine lengths with different spine length ranges in the latitude bands $0^{\circ}  \sim  50^{\circ}$  and $50^{\circ}  \sim  90^{\circ}$ in the northern and southern hemispheres from 1988 to 2013.} \label{table11}
\begin{tabular}{lclclclclc} \\
\hline \hline
&&Low latitude bands     &($0^{\circ}  \sim  50^{\circ}$)\\	
Time interval   		& Solar Cycle  &Spine length ($\times10^{4} $ km)		&$N_{C}$ (km)		&$S_{C}$ (km)			 &$A_{C}$    &Dominant hemisphere\\
\hline
1988 Apr -- 1996 May & 22   	     &$<5$ 				& $2.231\times10^{8}$ 		&$2.423\times10^{8}$       		 &$-0.041$ 		&South 		 \\
				&	  	     &$>5$   			& $4.043\times10^{8}$ 		&$5.176\times10^{8}$       		 &$-0.123$ 		&South		 \\
\hline
 1996 Jun -- 2008 Jan & 23 	     &$<5$	  			& $2.535\times10^{8}$    		&$2.479\times10^{8}$       		 &$0.011$ 			&North  		 \\
				&	  	     &$>5$   			& $1.107\times10^{9}$ 		&$1.007\times10^{9}$       		 &$0.047$ 			&North 		 \\
\hline
 2008 Feb -- 2013 Dec & 24             &$<5$                      	& $7.465\times10^{7}$           	&$6.602\times10^{7}$                   &$0.061$                	&North             \\
	                          &                    &$>5$           			& $2.607\times10^{8}$           	&$2.143\times10^{8}$                   &$0.098$                	&North            \\
\hline\hline
&&High latitude bands     &($50^{\circ}  \sim  90^{\circ}$)\\
Time interval   		& Solar Cycle  &Spine length ($\times10^{4} $ km)		&$N_{C}$ (km)		&$S_{C}$ (km)			 &$A_{C}$    &Dominant hemisphere\\
\hline
1988 Apr -- 1996 May & 22   	     &$<5$ 				& $1.269\times10^{7}$ 		&$1.028\times10^{7}$       		 &$0.105$ 			&North 		 \\
				&	  	     &$>5$   			& $2.182\times10^{7}$ 		&$1.873\times10^{7}$       		 &$0.076$ 			&North		 \\
\hline
 1996 Jun -- 2008 Jan & 23 	     &$<5$	  			& $1.412\times10^{7}$    		&$1.037\times10^{7}$       		 &$0.153$ 			&North  		 \\
				&	  	     &$>5$   			& $3.024\times10^{7}$ 		&$2.568\times10^{7}$       		 &$0.081$ 			&North 		 \\
\hline
 2008 Feb -- 2013 Dec & 24              &$<5$                      	& $3.376\times10^{6}$           	&$3.902\times10^{6}$                   &$-0.072$                	&South             \\
	                          &                     &$>5$           		& $3.762\times10^{6}$             	&$5.450\times10^{6}$                   &$-0.183$                	&South            \\	
\hline
\end{tabular}
\end{table*}

\begin{table*} \scriptsize
\centering
\caption{Asymmetry of the filaments with different tilt angle intervals in the latitude bands $0^{\circ}  \sim  50^{\circ}$  and $50^{\circ}  \sim  90^{\circ}$ in the northern and southern hemispheres from 1988 to 2013.} \label{table12}
\begin{tabular}{lclclclclclclclc} \\
\hline \hline
&&Low latitude bands     &($0^{\circ}  \sim  50^{\circ}$)\\	
Time interval   		& Solar Cycle  &Tilt angle interval				&$N_{N}$		&$S_{N}$			 &$A_{N}$    &Dominant hemisphere\\
\hline
1988 Apr -- 1996 May & 22   	     &$-90^{\circ}  \sim  -60^{\circ}$ 	& $680$ 		&$473$       		 &$0.180$ 		 	&North 		 \\
				&	  	     &$-60^{\circ}  \sim  -30^{\circ}$   	& $2,620$ 		&$1,652$       		 &$0.227$ 		 	&North		 \\
				&	  	     &$-30^{\circ}  \sim  0^{\circ}$   	& $3,489$ 		&$3,210$       		 &$0.042$ 			&North		 \\
				&	  	     &$0^{\circ}  \sim  30^{\circ}$  		& $2,566$ 		&$3,958$       		 &$-0.213$ 		&South		 \\
				&	  	     &$30^{\circ}  \sim  60^{\circ}$   	& $2,235$ 		&$3,673$       		 &$-0.243$ 		&South		 \\
				&	  	     &$60^{\circ}  \sim  90^{\circ}$   	& $444$ 		&$524$       		 &$-0.083$ 		&South		 \\
\hline
 1996 Jun -- 2008 Jan & 23 	     &$-90^{\circ}  \sim  -60^{\circ}$	& $1,336$    	&$795$       		 &$0.254$ 		 	&North  		 \\
				&	  	     &$-60^{\circ}  \sim  -30^{\circ}$   	& $4,164$ 		&$2,014$       		 &$0.348$ 		 	&North		 \\
				&	  	     &$-30^{\circ}  \sim  0^{\circ}$   	& $4,478$ 		&$2,790$       		 &$0.232$ 			&North		 \\
				&	  	     &$0^{\circ}  \sim  30^{\circ}$  		& $3,100$ 		&$4,297$       		 &$-0.162$ 		&South		 \\
				&	  	     &$30^{\circ}  \sim  60^{\circ}$   	& $2,036$ 		&$4,043$       		 &$-0.330$ 		&South		 \\
				&	  	     &$60^{\circ}  \sim  90^{\circ}$   	& $824$ 		&$1,193$       		 &$-0.183$ 		&South		 \\
\hline
 2008 Feb -- 2013 Dec & 24 	     &$-90^{\circ}  \sim  -60^{\circ}$  	& $342$   		&$203$      		 &$0.255$ 		 	&North  		 \\
				&	  	     &$-60^{\circ}  \sim  -30^{\circ}$   	& $1,158$ 		&$441$       		 &$0.448$ 		 	&Norh		 \\
				&	  	     &$-30^{\circ}  \sim  0^{\circ}$   	& $1,271$ 		&$668$       		 &$0.311$ 			&North		 \\
				&	  	     &$0^{\circ}  \sim  30^{\circ}$  		& $824$ 		&$980$       		 &$-0.087$ 		&South		 \\
				&	  	     &$30^{\circ}  \sim  60^{\circ}$   	& $497$ 		&$1,031$       		 &$-0.350$ 		&South		 \\
				&	  	     &$60^{\circ}  \sim  90^{\circ}$   	& $222$ 		&$346$       		 &$-0.218$ 		&South		 \\
\hline \hline
&&High latitude bands     &($50^{\circ}  \sim  90^{\circ}$)\\
Time interval   		& Solar Cycle  &Tilt angle interval				&$N_{N}$		&$S_{N}$			 &$A_{N}$	   &Dominant hemisphere\\
\hline
1988 Apr -- 1996 May & 22   	     &$-90^{\circ}  \sim  -60^{\circ}$ 	& $31$ 		&$19$       			 &$0.240$ 		 	&North 		 \\
				&	  	     &$-60^{\circ}  \sim  -30^{\circ}$   	& $57$ 		&$83$       		 	 &$-0.186$ 		&South		 \\
				&	  	     &$-30^{\circ}  \sim  0^{\circ}$   	& $160$ 		&$196$       		 &$-0.101$ 		&South		 \\
				&	  	     &$0^{\circ}  \sim  30^{\circ}$  		& $243$ 		&$165$       		 &$0.191$ 			&North		 \\
				&	  	     &$30^{\circ}  \sim  60^{\circ}$   	& $158$ 		&$89$       		 	 &$0.279$ 			&North		 \\
				&	  	     &$60^{\circ}  \sim  90^{\circ}$   	& $30$ 		&$16$       			 &$0.304$ 			&North		 \\
\hline
 1996 Jun -- 2008 Jan & 23 	     &$-90^{\circ}  \sim  -60^{\circ}$	 & $20$	    	&$26$       			 &$-0.130$ 		&South  		 \\
				&	  	     &$-60^{\circ}  \sim  -30^{\circ}$   	& $59$ 		&$126$       		 &$-0.362$ 		&South		 \\
				&	  	     &$-30^{\circ}  \sim  0^{\circ}$   	& $191$ 		&$190$       		 &$0.003$ 			&North		 \\
				&	  	     &$0^{\circ}  \sim  30^{\circ}$  		& $308$ 		&$136$       		 &$0.387$ 		 	&North		 \\
				&	  	     &$30^{\circ}  \sim  60^{\circ}$   	& $132$ 		&$47$       		 	 &$0.475$ 			&North		 \\
				&	  	     &$60^{\circ}  \sim  90^{\circ}$   	& $27$ 		&$24$       			 &$0.059$ 		 	&North		 \\
\hline
 2008 Feb -- 2013 Dec & 24 	     &$-90^{\circ}  \sim  -60^{\circ}$  	& $3$   		&$13$      		 	 &$-0.625$ 		 &South		 \\
				&	  	     &$-60^{\circ}  \sim  -30^{\circ}$   	& $14$ 		&$38$       			 &$-0.462$ 		 &South		 \\
				&	  	     &$-30^{\circ}  \sim  0^{\circ}$   	& $31$ 		&$67$       			 &$-0.367$ 		&South		 \\	
				&	  	     &$0^{\circ}  \sim  30^{\circ}$  		& $50$ 		&$29$       			 &$0.266$ 		 	&North		 \\
				&	  	     &$30^{\circ}  \sim  60^{\circ}$   	& $49$ 		&$17$       			 &$0.485$ 			&North		 \\
				&	  	     &$60^{\circ}  \sim  90^{\circ}$   	& $3$ 		&$8$       			 &$-0.455$ 		&South		 \\
\hline
\end{tabular}
\end{table*}

\clearpage
%figure 1
%figure 2

%figure 3

%figure 4

%figure 5

%figure 6

%figure 7

%figure 8

%figure 9

%figure 10

\clearpage

%figure11

%figure12

%figure13

%figure14

%figure15

%figure16

%figure17

%figure18

%figure19

%figure20

%figure21

%figure22

%figure23

%figure24

%figure25

\end{document}